\documentclass[11pt, letterpaper]{article}
\usepackage[english]{babel}
\usepackage[utf8]{inputenc}
\usepackage[margin=1in, bottom=1in, top=1in]{geometry}
\usepackage{fancyhdr}
\usepackage{graphicx, enumerate, amsmath, amsfonts, amssymb, amsthm, hyperref, braket, nicefrac}
\usepackage{listings}
\usepackage{array} 
\newcolumntype{x}[1]{>{\centering\arraybackslash\hspace{0pt}}p{#1}}
\usepackage{makecell} 
\usepackage{caption, subcaption, wrapfig, multirow}
\usepackage{booktabs}
\title{Temperature nonuniformity Part II}
\date{June 2020}

\DeclareUnicodeCharacter{03BC}{$\mu$}
\DeclareUnicodeCharacter{2212}{-}
\graphicspath{{snorm_figures/}}

\usepackage{xcolor} 

\begin{document}
{\Large
\textbf\newline{Resolving nonuniform temperature distributions with single-beam absorption spectroscopy. Part II: Implementation from broadband spectra}
}
\newline
\\
Nathan A Malarich\textsuperscript{1,*},
Gregory B Rieker\textsuperscript{1},
\\
\bigskip
1 Mechanical Engineering, University of Colorado at Boulder
\\
\bigskip
* nathan.malarich@colorado.edu

\begin{abstract}
Several past studies have described how absorption spectroscopy can be used to determine spatial temperature variations along the optical path by measuring the unique, nonlinear response to temperature of many molecular absorption transitions and performing an inversion.
New laser absorption spectroscopy techniques are well-suited to this nonuniformity measurement, yet present analysis approaches use only isolated features rather than a full broadband spectral measurement.
In this work, we develop a constrained spectral fitting technique called E$^{\prime\prime}$-binning to fit an absorption spectrum arising from a nonuniform environment.
The information extracted from E$^{\prime\prime}$-binning is then input to the inversion approach from the previous paper in this series (Malarich and Rieker, \textit{JQSRT} 107455 \cite{malarich_tx1}) to determine the temperature distribution.
We demonstrate this approach by using dual frequency comb laser measurements to resolve convection cells in a tube furnace. The recovered temperature distributions at each measurement height agree with an existing natural convection model.
Finally, we show that for real-world measurements with noise and absorption model error, increasing the bandwidth and the number of measured absorption transitions may improve the temperature distribution precision.
We make the fitting code publicly available for use with any broadband absorption measurement.
\end{abstract}

\textsuperscript{\textcopyright}2021. This manuscript version is made available under the CC-BY-NC-ND 4.0 license \\ http://creativecommons.org/licenses/by-nc-nd/4.0/.
This is the accepted version of the manuscript at https://doi.org/10.1015/j.jqsrt.2021.107805.

\section{Introduction}
Several scientific fields rely on single-path radiative transfer measurements to determine gas temperature and composition.
In combustion science, Laser absorption measurements can determine temperature \cite{twoline_first, sanders_2001, lbin, ma_tomographic_2009}, as well as concentrations of reactive intermediates \cite{hayden_oh} and pollutant products \cite{langridge_cavity_2008, chao_development_2013}.
In atmospheric science, the temperature profile and greenhouse gas concentrations of the Earth atmosphere are determined with passive ground-based instruments \cite{tccon, schneider_ftir_profile}] and downward-facing satellites \cite{oco2_inversion, oco2_cities}.
In astronomy, transmission spectra through exoplanet atmospheres propagate to Earth's space- and ground-based telescopes, providing evidence of exoplanet stratospheres \cite{knutson_stratosphere, wasp_stratosphere, evans_stratosphere} and molecular constituents \cite{exoplanet_sodium_d, exoplanet_bayes}.

Each of these applications measure the transmission spectrum of radiation across the gas system, from which post-processing algorithms calculate temperature and composition. Nonuniformity in temperature, composition, and pressure across the optical path alters these measured transmission spectra.
These path-nonuniformity alterations introduce uncertainty into average-property temperature and composition measurements \cite{goldenstein_nonuni}.
However, certain transmission spectra may contain enough information to determine line-of-sight nonuniformity based on how the transmission spectra are distorted \cite{sanders_2001}. 

Thermodynamic nonuniformity influences the infrared transmission measurement at several different spectral scales, and each scale can be leveraged in a unique way to infer nonuniformity.
At the coarsest spectral scale of $\sim$10-100 cm\textsuperscript{-1}, molecular emission from gases over optically thick paths can dominate the characteristics of the transmission spectrum.
This scale is used to determine vertical temperature profiles of planetary atmospheres, by fitting Planck blackbody curves to several regions of an emission spectrum at different opacities. \cite{knutson_stratosphere, wasp_stratosphere, evans_stratosphere, petty_atmospheric_2006}.
At an intermediate scale of $\sim$0.1-1 cm\textsuperscript{-1}, the magnitudes of individual quantum absorption features in gases each have a different nonlinear temperature-dependence described by the Boltzmann distribution (Eq. \ref{eq:boltz}), which can be used to determine temperature variability, for example in combustion systems \cite{malarich_tx1, sanders_2001, lbin, ma_tomographic_2009, beijing_nonuni}.
\begin{equation}
\label{eq:boltz}
\textrm{absorption feature integrated area} \sim exp\left(-\frac{c_2 E^{\prime\prime}}{T}\right)
\end{equation}
Here $c_2$ is the product of several physical constants, $T$ is the gas temperature, and E$^{\prime\prime}$ is the lower-state energy of a particular absorption feature.
By measuring the integrated areas of many features with different lower-state energies, one can determine the relative populations of molecules in different rotational-vibrational quantum states throughout a laser path.
The populations can then be linked to the collection of different Boltzmann distributions induced by temperature nonuniformity along a laser path.
This approach determines a path-integrated quantity of gas molecules at the temperature associated with each Boltzmann distribution -- a position-insensitive ``temperature distribution'' \cite{malarich_tx1}.
At the smallest scale of spectral resolution is the lineshape of the individual absorption transitions.
To determine vertical concentration profiles from a single atmospheric transmission measurement, one can distinguish the broader absorption signature of the high-pressure troposphere from the narrower absorption of the low-pressure upper atmosphere.
This method requires $<$0.1 cm\textsuperscript{-1} resolution in the transmission spectrum and pressure variation along the path \cite{gfit2, tccon_profile, ramanathan_svd}. 

This paper uses the intermediate spectral scale -- leveraging the magnitude (integrated absorbance area) of a large number of quantum absorption features to recover the temperature distribution along the optical path.
This integrated area approach produces the strongest temperature nonuniformity signals in the transmission spectrum for optically thin systems with similar pressure across the measurement path (e.g. typical to combustion diagnostics).
The low-resolution blackbody emission is insignificant for laser absorption measurements of optically thin systems \cite{hanson_reader}, and the high-resolution lineshape approach is ineffective for gases with relatively uniform pressure (whereas the vertical atmospheric applications have orders of magnitude pressure variation along the path).

\begin{figure}
 \centering
  \includegraphics{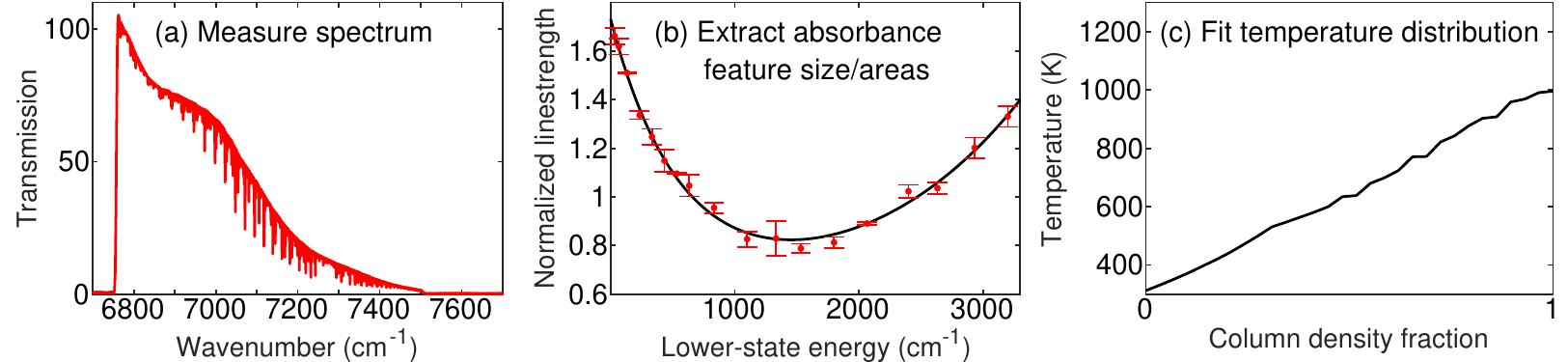}
  \caption{A nonuniform temperature measurement using absorption spectroscopy is a three-step process. The companion paper \cite{malarich_tx1} describes step 3 -- how to determine the temperature distribution corresponding to a curve-fit of normalized linestrengths (which are proportional to the integrated areas of absorption features). This paper shows how to extract normalized linestrengths from a transmission spectrum (step 2), then combines the three steps in a validation experiment.}
\label{fig:3step}
\end{figure}

We describe our temperature nonuniformity measurement approach through the progression in Fig. \ref{fig:3step}.
The transmission specrum $I(\nu)$ of Fig. \ref{fig:3step}a is the measurement from the broadband spectrometer that encodes the temperature distribution.
The temperature distribution information is contained in the relative areas of the individual absorption features.
This paper will describe an efficient means to fit the integrated areas of the absorption features.
We manipulate the integrated areas into ``normalized linestrengths,” which describe the normalized laser-path-integrated population of molecules in quantum state of particular lower-state energies (the red points in Fig. \ref{fig:3step}b) \cite{malarich_tx1}.
Simulating this normalized linestrength curve requires fewer parameters than simulating the full transmission spectrum,  thus enabling an efficient means to solve for the temperature distribution.
The temperature distribution solution shown in Fig. \ref{fig:3step}c is a monotonically-increasing function which describes the position-insensitive fraction of absorbing molecules in the laser path below each particular temperature.
We note the similarity of this three-step procedure to other metrology techniques, which first extract concentrations from transmission spectra and then use several such concentrations to quantify a spatial characteristic of the gas \cite{grauer_jqsrt, longpath_optica}.

The Part I companion to this paper \cite{malarich_tx1} studied the theoretical capabilities of the final step of the technique: fitting for a temperature distribution (Fig. \ref{fig:3step}c) from a set of normalized linestrengths in Fig. \ref{fig:3step}b.
This analysis omitted the complexity of the absorption model parameters that is present in the full spectrum-to-temperature technique.
The companion paper defined a matrix (adapted from \cite{sanders_2001}) which calculates a particular normalized linestrength curve (Fig. \ref{fig:3step}b black trace) from a given temperature distribution (Fig. \ref{fig:3step}c black trace).
One can invert this matrix to solve for the most likely temperature distribution given a vector of normalized linestrengths (Fig. \ref{fig:3step}b red points).
The companion paper found that while the technique can recover several temperature distributions within 3\% accuracy, the Step 3 matrix inversion is ill-posed.
In other words, one must use caution, as several different monotonic temperature distribution curves in Fig. \ref{fig:3step}c can produce identical (i.e. to within machine precision) normalized linestrength curves in Fig. \ref{fig:3step}b.

Interestingly, the companion paper also showed that in a measurement of normalized linestrengths to machine precision, just $\sim$15 data points on Fig. \ref{fig:3step}b produce the same temperature distribution solution as an equally perfect measurement of the normalized linestrengths of hundreds of absorption features over the same range of lower-state energy.
However, any experimental measurement of an absorption feature will necessarily introduce uncertainty into the normalized linestrength (e.g. through noise), which increases the resulting temperature distribution uncertainty and complicates this 15-feature finding.
Therefore, the accuracy of the temperature distribution solution relies on both the number of absorption features and the ability to accurately determine the normalized linestrengths of those features from the spectrum.

Broadband, high-resolution spectrometers that measure many more than 15 absorption features may better determine the temperature distributions due to the multiplicity of features.
By the standard error relationship, measuring more absorption features should make a result more robust against a variety of random components of measurement and model errors present in real spectroscopy measurements.
The challenge then is to leverage the breadth of absorption features to extract accurate integrated areas, while retaining sensitivity to the temperature-nonuniformity and not burdening the data processing infrastructure with a large number of fit parameters.
This paper introduces a constrained spectral fitting method based on the normalized linestrength to accomplish this task, leveraging concepts from broadband fitting of multiple species in absorption spectroscopy \cite{werblinski_temperature_2013, diba_multispecies, hundt_multispecies, kara_multispecies, jahromi_multispecies}.
A code package to perform this fitting method is also available with this paper \cite{ntfit}.

We demonstrate an entire nonuniform temperature measurement from broadband absorption spectroscopy, containing all three steps depicted in Fig. \ref{fig:3step}.
In Section \ref{sec:ebin}, we introduce the constrained spectral fitting method, named E$^{\prime\prime}$-binning, in order to extract normalized linestrengths (Fig. \ref{fig:3step}b) from a transmission spectrum (Fig. \ref{fig:3step}a).
In Section \ref{sec:expt}, we experimentally demonstrate the full nonuniform-temperature technique in an isobaric, uniform-concentration, nonuniform-temperature system, combining the E$^{\prime\prime}$-binning spectral fit from Section \ref{sec:ebin} and the temperature inversion step from \cite{malarich_tx1} with uncertainty estimation. We extract temperature distributions from dual frequency comb laser measurements in a tube furnace, which agree within experimental uncertainty to a natural convection model.
Last, in Section \ref{sec:bandwidth}, we show why using broadband measurements with more than 15 absorption features improves the temperature distribution solution accuracy.

\section{E$^{\prime\prime}$-binning: extracting linestrengths from transmission spectrum}
\label{sec:ebin}
We present a new spectral fitting method that is particularly useful for determining line-of-sight temperature nonuniformity from a broadband transmission spectrum.
This method becomes Step 2 of the 3-step temperature inversion outlined in Fig. \ref{fig:3step}.
In this section, we first motivate this method, then simulate its performance in the context of the 3-step inversion, and finally discuss refinements of this method.
We have released a code package, NTfit, which implements the E$^{\prime\prime}$-binning method described in this section.  The code package is available at https://gitlab.com/nam5312/ntfit \cite{ntfit}.

\subsection{Background}
In order to minimize temperature distribution uncertainty, one needs to measure the precise integrated area of at least 15 absorption features over a wide range of lower-state energy (at least 3000 cm\textsuperscript{-1}) \cite{malarich_tx1}.
To fulfill this criterion for near-infrared water vapor features, one must measure features spanning hundreds of wavenumbers.
The broadband laser sources (such as dual frequency combs) that are well-suited to this task will simultaneously measure hundreds of additional absorption features.
To reduce temperature distribution uncertainty from a spectrum with noise, it is beneficial to fit all of the measured absorption features.
Methods such as Voigt fitting \cite{velocity_nonuni} and hybrid-Voigt fitting \cite{sanders_2001} use several fit parameters for each absorption feature.
To expand such line-by-line methods across all the features in a broadband spectrum requires hundreds of degrees of freedom,
and assigning the integrated area to the appropriate feature in congested regions of the spectrum is challenging.
Here, we offer an alternative approach with far fewer fit parameters to extract hundreds of accurate integrated areas from a broadband spectrum.

Rather than fit individual features, some methods fit absorption signatures for the full broadband spectrum.
These methods typically simulate a multidimensional matrix of theoretical spectra at a variety of temperatures ($T$), pressures ($P$), and species molefractions ($\chi_i$), and then perform a gradient-based nonlinear least-squares fitting method to fit the path-averaged $T$, $P$, $\chi_i$ to a measured spectrum \cite{rieker_outdoor, werblinski_supercontinuum, kranendonk_robust}. 
Recent work has built on these broadband-fitting approaches, using a modified form of the molecular free induction decay (m-FID) signal to extract gas properties without needing to fit the baseline laser intensity spectrum \cite{time_domain}.
We present an adaptation of this full-spectrum baseline-free algorithm that captures the nonuniform-temperature spectral signal, and quickly produces reasonable estimates of the integrated areas using a linear least-squares fitting method.

\subsection{Fitting normalized linestrengths}
The normalized linestrength defined in the companion paper (Section 2.2 of \cite{malarich_tx1}) and below presents a foundation for our method of constrained spectral fitting from a nonuniform environment.
From this context, the goal of the spectral fitting method is to extract normalized linestrengths from the transmission spectrum to create the normalized linestrength versus lower state energy curve in Fig. \ref{fig:3step}b to high accuracy.
This normalized linestrength curve can then be fit to determine the temperature distribution \cite{malarich_tx1}. 

The integrated area ($\int_\nu \alpha_i$) of each absorption feature is traditionally calculated with a laser-path integral. The companion paper \cite{malarich_tx1} instead showed how to express this integrated area in terms of an order-insensitive temperature distribution function (TDF) and linestrength ($S$):
\begin{equation}
\label{eq:dT}
\int_\nu \alpha_i = \overline{P \chi} L \int_0^\infty S_i(T_j) TDF(T_j) dT_j
\end{equation}

The companion paper showed how to solve Eq. \ref{eq:dT} for the TDF and to produce the temperature plots of Fig. \ref{fig:3step}c from the TDF \cite{malarich_tx1}.
Each absorption feature, $i$,  has a known temperature-dependent linestrength function $S_i(T)$ with a particular lower-state energy E$_i^{\prime\prime}$.
 The companion paper showed how to produce a matrix from these linestrengths $S_i(T_j)$, where each row represents a different absorption feature measurement ($\int_\nu \alpha_i$) and each column represents the unknown path-integrated fraction of light-absorbing molecules at a particular temperature $T_j$ \cite{malarich_tx1}.

Rather than fit a unique integrated area for each absorption feature as in the hybrid Voigt method, we can instead fit a single normalized linestrength for many absorption features with similar lower state energies.
We define the normalized linestrength in Eq. \ref{eq:snorm} by dividing both sides of Eq. \ref{eq:dT} by the linestrength of the feature $S_i(T_0)$ at some reference temperature $T_0$.
Two absorption features with the same lower-state energy E$^{\prime\prime}$ will typically have different integrated areas, but barring absorption model error or stimulated emission differences (e.g. for high-temperature mid-IR spectra) should always have the same normalized linestrength.
We will use the left-hand side of Eq. \ref{eq:snorm} to extract linestrengths from the spectrum, and the companion paper used the right-hand side of Eq. \ref{eq:snorm} to fit a temperature distribution to a set of linestrengths.
Terms with subscript $i$ have implicit E$_i^{\prime\prime}$ dependence.
This approach is particularly helpful for complex spectra with many weak and overlapping features of unique temperature-dependence, where the hybrid Voigt method is challenging.

\begin{equation}
\label{eq:snorm}
\hat{S_i}(E_i^{\prime\prime};T_0) = \underbrace{\frac {\int_\nu \alpha_i} {S_i(T_0)}}_\textrm{Derived from $\int\alpha$ measurement}  = \underbrace{\overline{P \chi}L \int_0^\infty \frac{S_i(T_j)}{S_i(T_0)} TDF(T_j)dT_j}_\textrm{Calculated from TDF}
\end{equation}

\begin{figure}[h]
 \centering
 \includegraphics{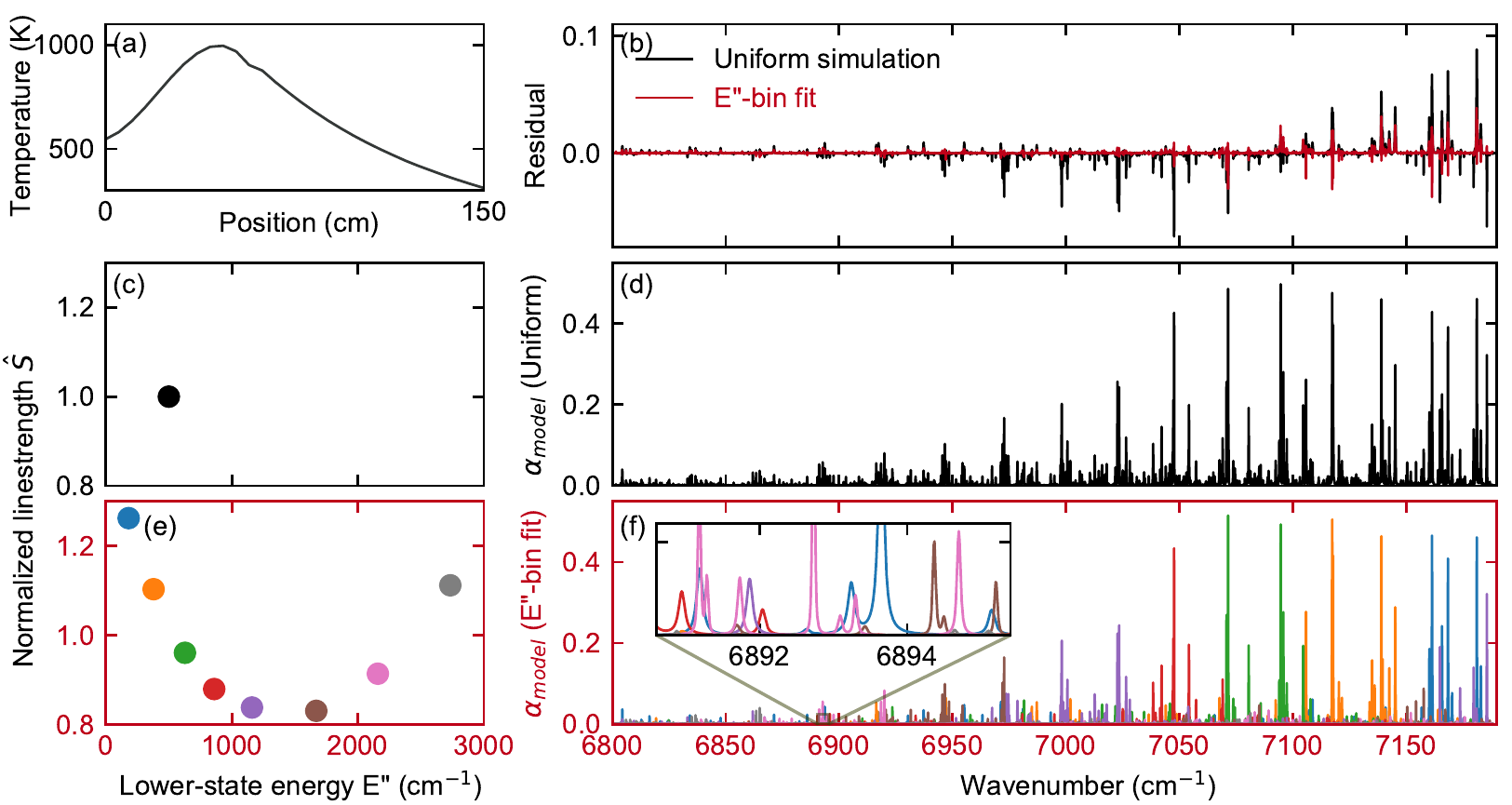}
  \caption{E$^{\prime\prime}$-binning approach demonstrated on the nonuniform temperature profile shown on (a). (b) Spectral residual ($\alpha\textsubscript{meas} – \alpha_{model}$) between nonuniform-temperature simulation spectrum (not shown) and two models. Absorbance residual is smaller for E$^{\prime\prime}$-bin fit. (d) Uniform-temperature simulation $T$, $P_{H_2O}$ with (c) corresponding normalized linestrength plot.
(f) More complex fit of 8 E$^{\prime\prime}$-bins. Each color represents different E$^{\prime\prime}$-bin group of H\textsubscript{2}O absorption features.
Inset shows many overlapping absorption features with different temperature-dependence.
(e) Normalized linestrength plot. Y-axis shows normalized linestrength fitted for each bin of all H\textsubscript{2}O absorption features with similar lower-state energy.}
\label{fig:ebin}
\end{figure}

Figure \ref{fig:ebin} shows how the E$^{\prime\prime}$-binning technique works for a water vapor absorption spectrum.
We calculate a synthetic ``measured" absorption spectrum ($\alpha\textsubscript{meas}$) from the nonuniform temperature profile in Fig. \ref{fig:ebin}a, and then compare this spectrum with two absorption models: a uniform-path simulation at the average thermodynamic conditions, and a fit using the E$^{\prime\prime}$-binning technique.
The uniform-path simulation is shown in Fig. \ref{fig:ebin}d.
The corresponding ``uniform simulation” residual in Fig. \ref{fig:ebin}b encodes the information about temperature nonuniformity.
This ``uniform” spectrum forces every H\textsubscript{2}O absorption feature to have the same integrated area as predicted by a single $T$, $P$, $\chi$, $L$, even though the nonuniform-temperature TDF in the Eq. \ref{eq:snorm} integral produces different relative integrated areas for different absorption features in $\alpha\textsubscript{meas}$.
In Fig. \ref{fig:ebin}c, this uniform model is analogous to forcing every absorption feature to have the same unity value of the normalized linestrength, regardless of the feature's E$^{\prime\prime}$.
Fig. \ref{fig:ebin}c is shown with a single point representing the weighted-average E$^{\prime\prime}$ for the spectrum.

In the E$^{\prime\prime}$-binning approach, we split the H\textsubscript{2}O features into different `bins’ according to their E$^{\prime\prime}$, and fit for the normalized linestrength of each bin separately.
This approach reduces the spectral residual shown in Fig. \ref{fig:ebin}b.
E$^{\prime\prime}$-binning begins by simulating a set of absorption models, $\alpha_{0,i}(\nu)$, each containing all of the H\textsubscript{2}O features within a particular range of E$^{\prime\prime}$.
Then we determine a set of linear scaling coefficients, $\hat{S}_i$, to fit the measured absorbance spectrum, as shown in Eq. \ref{eq:snorm_fit}.

\begin{equation}
\label{eq:snorm_fit}
\min_{\hat{S}_i} \left \lVert \alpha\textsubscript{meas}(\nu) - \sum\limits_{i=1}^{\# E^{\prime\prime} \ bins} \hat{S}_i \ \alpha_{0,i}(\nu)   \right \rVert
\end{equation}

The absorption model ($\alpha_{0,i}(\nu)$) is the predicted absorbance signal due to all of the relevant H\textsubscript{2}O features at some nominal $T_0$, $P_0$, $\chi_0$, $L_0$.
The linear scaling coefficients (normalized linestrength) are the average integrated area of the absorption features in that E$^{\prime\prime}$-bin normalized by the predicted integrated area at a nominal condition.
We plot these normalized linestrengths on Fig. \ref{fig:ebin}e with respect to the average E$^{\prime\prime}$ of each set of features.

This E$^{\prime\prime}$-binning fit is analogous to multi-species fitting in broadband absorption spectroscopy \cite{makowiecki_mid-ir}.
In multi-species fitting, each species has an absorption model, and the linear fit coefficient that is multiplied by each model represents the species concentration.
In E$^{\prime\prime}$-binning, one species is split into several distinct absorption models (of all the absorption features with similar E$^{\prime\prime}$), and the fit coefficients for each signature are the normalized linestrengths at each E$^{\prime\prime}$ bin.
Additionally, we can use modified free induction decay (m-FID) fitting \cite{time_domain} to separate the laser intensity baseline from most of the absorption signal, and determine the normalized linestrengths by linear regression (see Appendix \ref{sec:mfid}).
We can then use the inversion method in the companion paper to solve for a temperature distribution that fits a convex curve to the normalized linestrengths.

Fig. \ref{fig:ebin}e shows normalized linestrengths greater than unity for the E$^{\prime\prime}$$<$500 cm\textsuperscript{-1} bins. This means that the features with E$^{\prime\prime}$$<$500 cm\textsuperscript{-1} are larger than anticipated at the nominal $T_0$, $P_0$, $\chi_0$, $L_0$.
Physically, this means that more water molecules are in the lower-energy range of rotational-vibrational quantum states than one would expect from a Boltzmann distribution at temperature $T_0$, because a fraction of the path is at lower temperatures than $T_0$.
Figure \ref{fig:ebin}e also shows that the E$^{\prime\prime}$$>$2500 cm\textsuperscript{-1} bin has normalized linestrength greater than unity, suggesting that a fraction of the path lies above temperature $T_0$.
The intermediate-range E$^{\prime\prime}$ bins have the smallest normalized linestrengths in Fig. \ref{fig:ebin}e, indicating that there is temperature nonuniformity; if all of the E$^{\prime\prime}$-bins had normalized linestrengths above unity, that would simply indicate that the column density estimate $P_0 \chi_0 L_0$ was too low.
The method described in the companion paper \cite{malarich_tx1} takes the set of normalized linestrengths $\hat{S}(E^{\prime\prime})$, the nominal temperature $T_0$, and the column density $P_0 \chi_0 L_0$ to determine the temperature distribution, an order-insensitive version of the profile in Fig. \ref{fig:ebin}a.

There are two advantages to fitting normalized linestrengths rather than individual integrated areas of each feature.
First, the normalized linestrength represents the integrated areas of many absorption features with a single degree of freedom.
By reducing the degrees of freedom compared to a line-by-line fitting method, the E$^{\prime\prime}$-binning method can more robustly fit complex portions of the spectra such as the portion shown in the inset of Fig. \ref{fig:ebin}f, where there are many overlapping absorption features with unique temperature-dependence. 
A single normalized linestrength for many absorption features should have a higher precision than a single integrated area.
We discuss this precision improvements from such full-spectrum fitting in Section \ref{sec:bandwidth}.
Second, the normalized linestrength curve in Fig. \ref{fig:ebin}e offers a visual of the fit quality. The normalized linestrengths should form the shape of a curve with small scatter and uncertainty bars. We discuss the utility of this visual in Section \ref{sec:expt2}.

\subsection{Demonstration on noiseless synthetic spectra}
\label{sec:noiseless}
\begin{figure}[h]
 \centering
  \includegraphics{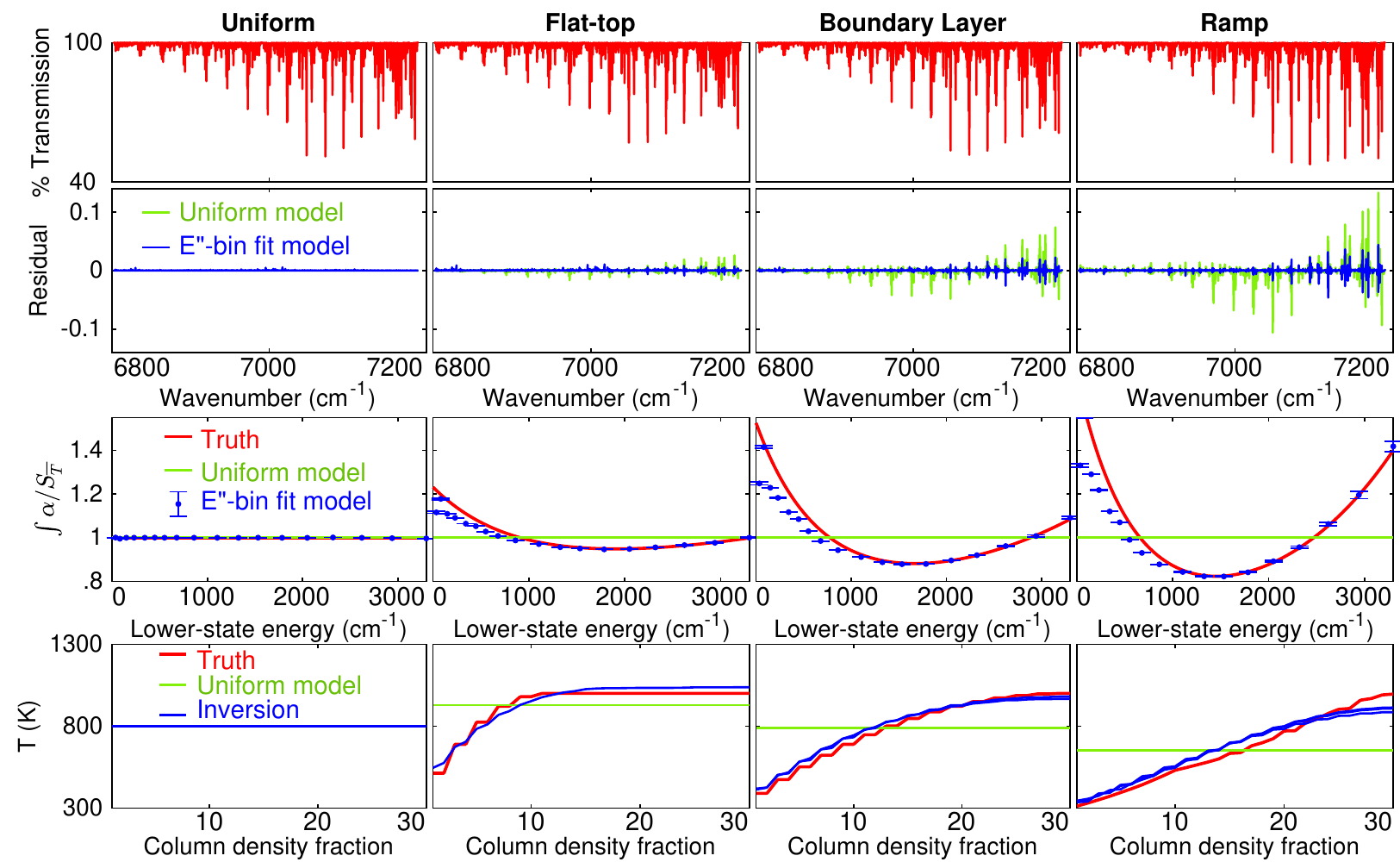}
  \caption{Temperature inversion algorithm applied to noiseless synthetic spectra for four temperature profiles (1\% H\textsubscript{2}O in 1 atm air). (Left-to-right) different nonuniform temperature profiles. (top) Synthetic transmission spectra. (2\textsuperscript{nd} from top) Spectral fitting residual due to fitting the transmission spectrum from (top) with the absorption models. (3\textsuperscript{rd} from top) normalized linestrength fits. (bottom) Temperature distributions. Truth shown in red, nonuniform-temperature fit results shown in blue, and average-temperature broadband model shown in light green.}
\label{fig:noiseless}
\end{figure}

We first demonstrate this E$^{\prime\prime}$-binning method and the subsequent temperature distribution inversion on synthetic noiseless spectra from four different temperature profiles.
Figure \ref{fig:noiseless}(bottom) shows the four temperature profiles, and Fig. \ref{fig:noiseless}(top) shows the corresponding simulated spectra for each profile.
We evaluate the spectra with both uniform and E$^{\prime\prime}$-bin methods, producing the spectral residuals in the 2\textsuperscript{nd} row of plots and the normalized linestrength coefficients plotted in the 3\textsuperscript{rd} row.
The ``truth” in the normalized linestrength plot is calculated from Eq. \ref{eq:dT} for the temperature profiles, whereas the E$^{\prime\prime}$-bin fit comes from the absorption spectrum fit in Eq. \ref{eq:snorm_fit}.
We use the Tikhonov regularization method from the companion paper \cite{malarich_tx1} to determine the temperature distribution from the normalized linestrengths (Fig. \ref{fig:noiseless}(bottom)).
The light green traces represent the uniform-temperature simulation; the difference between this simulation and the truth traces in red represents the effect of temperature nonuniformity upon the transmission spectrum, normalized linestrengths, and temperature distribution, respectively. 

The E$^{\prime\prime}$-binning technique recovers a near-correct temperature distribution for all four cases.
However, the technique deviates from the truth at each step of the algorithm -- the absorbance residuals from the E$^{\prime\prime}$-bin fits are nonzero, the normalized linestrengths do not perfectly track the truth curve, and the temperature distributions from the inversion also deviate from the truth curves.
These deviations occur even in the absence of measurement noise or absorption model error, because there are several potential error sources in the E$^{\prime\prime}$-binning fit algorithm, as shown in Table \ref{tab:errors}.
In practice, these fitting algorithm error sources are either small, or we can modify this E$^{\prime\prime}$-binning algorithm to mitigate the error. The rest of Section \ref{sec:ebin} steps through these error sources and mitigation techniques.

\begin{table}
\begin{small}
\renewcommand\arraystretch{1.2}
\begin{tabular}{p{1.3mm} | p{5.1cm} p{7.8cm} p{1.8cm}}
 \toprule
\multicolumn{2}{c}{Source of linestrength uncertainty} & \multicolumn{2}{r}{Our mitigation technique        \text{ } \text{ }\text{ }\text{ }\text{ }\text{ } Improves with bandwidth} \\
\midrule
\multirow{7}{*}{\rotatebox[origin=c]{90}{Fitting algorithm}}
& 1) Spectral parameters & Use empirical 300-1300 K database & \multicolumn{1}{c}{Yes} \\
& 2) Incorrect average $P,\chi ,T$ guess  & Iterate average $P, \chi ,T$ from inversion  & \multicolumn{1}{c}{Yes} \\
& 3) $P,\chi ,T$ nonuniformity  & Fit broadening \& shift of each E$^{\prime\prime}$-bin  & \multicolumn{1}{c}{No} \\
& 4) Baseline determination and  & Group overlapping features by E$^{\prime\prime}$-bin  & \multicolumn{1}{c}{Yes} \\
& \text{ }\text{ } overlapping absorption features & & \\
& 5) Discretization error:  & Keep strongest features in separate E$^{\prime\prime}$-bins  & \multicolumn{1}{c}{No}\\
& \text{ } \text{ }$\hat{S}$ gradients across E$^{\prime\prime}$-bin & & \\
& 6) Stimulated emission & Negligible in near-infrared & \multicolumn{1}{c}{No} \\
& \text{ } \text{ }differences across E$^{\prime\prime}$-bin & & \\
\addlinespace[3mm]
\multirow{4}{*}{\rotatebox[origin=c]{90}{Experiment}}
& 7) Etalon + transmission noise & Time-average. Time-domain filter etalons & \multicolumn{1}{c}{Yes} \\
& 8) Background absorption  & Dry-air purge w/ RH sensors + bkgd measurement & \multicolumn{1}{c}{No} \\
& 9) Simulation/experiment & Self-referenced dual-frequency-comb spectrometer & \multicolumn{1}{c}{No} \\
& \text{  }\text{ } wavelength matching& & \\
\bottomrule
\end{tabular}
\end{small}
\caption{Potential sources of uncertainty in nonuniform-temperature retrievals with absorption spectroscopy}
\label{tab:errors}
\end{table}

The purpose of the E$^{\prime\prime}$-bin fitting algorithm is to extract the correct integrated areas, reducing interference from both broad bandwidth scales (e.g. baseline) and narrow bandwidth scales (e.g. lineshape).
The fitting algorithm errors in Table \ref{tab:errors} describe ways that these different bandwidth scales can influence the integrated area fits.
Baseline intensity is constant across the spectrum in this ideal synthetic case, while for a true measurement the baseline must be fit with additional parameters or filtered using m-FID.
Lineshape models are particularly important to fit spectral regions with overlapping absorption features; we will show in Section \ref{sec:expt2} that these overlapping-feature regions are prevalent throughout near-infrared H\textsubscript{2}O spectra.
The advantage of E$^{\prime\prime}$-binning is that it can still fit these spectral regions, by constraining the spectral model in the congested region to be consistent with other isolated features elsewhere in the spectrum.
Lineshape error can also impair the fitting algorithm through spectral parameter errors and $P$, $\chi$, $T$ effects (sources 1-3 in Table \ref{tab:errors}), which we discuss in Section \ref{sec:lineshape}.

In addition to baseline and lineshape interference, the E$^{\prime\prime}$-binning technique may recover an incorrect normalized linestrength due to spectral parameter error and discretization error.
Spectral parameter error in the reference linestrength ($S_{296}$ using the nomenclature from the primary HITRAN database reference \cite{hitran2016}) may shift the normalized linestrength, because a correct integrated area divided by an incorrect reference linestrength produces an incorrect normalized linestrength.
There is no spectral parameter error in the Fig. \ref{fig:noiseless} simulation.

Discretization error occurs because the fit requires all integrated areas with similar E$^{\prime\prime}$ to have the same normalized linestrength, when in reality there is slight variation because a particular E$^{\prime\prime}$-bin covers a range of E$^{\prime\prime}$.
This discretization error can be reduced by increasing the number of E$^{\prime\prime}$-bins.
However, adding E$^{\prime\prime}$-bins increases the number of fit parameters, and using large numbers of E$^{\prime\prime}$-bins makes the fitting algorithm more susceptible to instability from two other error sources: transmission noise and overlapping absorption features.
We have found for this wavelength region that providing a different E$^{\prime\prime}$-bin for each rotational quantum number in the strongest quantum transition family ($J^{\prime\prime} = K_c$, $\nu_1$+ $\nu_3$ band) provides a good compromise point for this tradeoff between stability and accuracy.
In section \ref{sec:expt2} we show justification for this particular E$^{\prime\prime}$-bin selection based on experimental data.

\subsection{Reducing lineshape error}
\label{sec:lineshape}
The largest source of error in the normalized linestrength plots of Fig. \ref{fig:noiseless} arises from error in the integrated area retrievals due to nonuniformity-induced lineshape error in the absorption models.
Lineshapes are calculated from spectral absorption parameters together with thermodynamic parameters $T$, $P$, $\chi$.
Assuming a single $T_0$, $P_0$, $\chi_0$ across all E$^{\prime\prime}$-bins will cause error in the fit to data from nonuniform environments.
We will show how to reduce lineshape error by adding a fourth step to the E$^{\prime\prime}$-binning algorithm as diagrammed in Table \ref{tab:alg}.

\begin{table}
\renewcommand{\arraystretch}{1.2}
\begin{small}
\begin{tabular}{p{3.9cm} p{2.3cm} p{5.2cm} p{3.4cm}}
 \toprule
Description &  \makecell{Fit \\parameters} & \makecell{Mitigated uncertainty (UC) \\ (sources from Table \ref{tab:errors})} & \makecell{NTfit\\ function} \\
 \midrule 
\makecell[tl]{1) Guess path-average \\ \text{ }\text{ }\text{ }$T_0, P_0, \chi_0, L$}  & \multicolumn{1}{c}{N/A}  & 
      \makecell[tl]{Incorrect path-average inputs \\ (Table 1, UC (2))} & EbinHapi.print\_thermo \\
\addlinespace[0.2cm]

2) Select edges of E$^{\prime\prime}$ bins &  \multicolumn{1}{c}{N/A}  & \makecell[tl]{Discretization error \\ (Table 1, UC (5))} & ``.Elist \\
\addlinespace[0.2cm]

3) m-FID fit    & \makecell{$\sim$15 $\hat{S}$}  & Baseline, overlapping absorption features (Table 1, UC (4))& ``.fit\_snorm\\
\addlinespace[0.2cm]

4) Update path-average $T_0$ and repeat fit & \makecell{$T_0$, $\chi_0$} &  \makecell[tl]{Incorrect path-average inputs \\ (Table 1, UC (2))} & ``.fit\_temperature \\

\makecell[tl]{5) Repeat fit \\ $\ \ \ $ with lineshape fit}   & 
	\vline \makecell[tl]{$\ \sim$15 $\hat{S}$ \\
					\text{ }$\sim$15 widths $\Gamma$ \\
					\text{ }$\sim$15 shifts $\Delta$}   & 
\makecell[tl]{$P,\chi ,T$ nonuniformity \\ (Table 1, UC (3)) (see Fig. \ref{fig:lineshape})} & ``.fit\_snorm\_width\_shift \\
\addlinespace[0.2cm]
 \bottomrule
\end{tabular}
\end{small}
\caption{E$^{\prime\prime}$-binning spectral fitting algorithm to extract linestrengths from a transmission spectrum. Includes corresponding function names from NTfit package \cite{ntfit} and uncertainties from Table \ref{tab:errors}.}
\label{tab:alg}
\end{table}

We use the versatile Voigt lineshape for the E$^{\prime\prime}$-bin fit.
The Voigt profile has three shape parameters in addition to integrated area: Doppler width, Lorentz width, and linecenter (with pressure-shift).
Each shape parameter includes temperature dependence, which can be described with a power-law (shown for Lorentz-width, $\Gamma$, in Eq. \ref{eq:lineshape}).
The widths and pressure shift scale from their reference-temperature values ($\Gamma_{T_0}$) to their observed values by $T^{-n}$.  

\begin{equation}
\label{eq:lineshape}
\Gamma = \Gamma_{T_0} \left(\frac{T_0} {T}  \right)^n
\end{equation}

Temperature nonuniformity influences the path-average Voigt parameters through this nonlinear power-law temperature dependence.
Calculating lineshapes even with the true path-averaged $T$, $P$, $\chi$ may thus produce incorrect linewidths and linecenter shifts.
Voigt parameter error is most pronounced for the low-E$^{\prime\prime}$ bins, because most of the absorption for low-E$^{\prime\prime}$ features occurs at locations along the path that are below the path-average temperature.
For example, in the Ramp test case (Fig. \ref{fig:noiseless}(right)), the E$^{\prime\prime}$=100 cm\textsuperscript{-1} features have very strong absorption below the path-average temperature of 653 Kelvin, and almost no absorption above that temperature.
The temperature-dependence exponent, $n$, is typically between 0.5-1, so the widths and shifts decrease at higher temperatures.
Therefore, the absorption model, $\alpha_{0,E^{\prime\prime}=100}(\nu; T=653)$, which is calculated at the path-average temperature for the Fig. \ref{fig:noiseless} fit, contains overly narrow absorption features for the E$^{\prime\prime}$=100 cm\textsuperscript{-1} bin. Without the correct lineshape model, the spectral fit cannot effectively determine the integrated areas, so the retrieved normalized linestrengths are incorrect.
In a nonuniform-temperature environment, we expect the Lorentz width modeled at the average temperature and pressure to be too small for the lowest-E$^{\prime\prime}$ bins, and too large for the highest-E$^{\prime\prime}$ bins. Similarly, H\textsubscript{2}O pressure-shift has a negative room-temperature coefficient with a positive power-law temperature-dependence exponent, so we expect the linecenter positions modeled at the average temperature and pressure to be too high for the lowest-E$^{\prime\prime}$ bins, and too low for the highest-E$^{\prime\prime}$ bins.

In order to compensate for this systematic E$^{\prime\prime}$-dependent Voigt parameter error that is induced by modeling at the average temperature and pressure, we add two degrees of freedom to each E$^{\prime\prime}$-bin. We fit a different pressure for each E$^{\prime\prime}$-bin to compensate for the Lorentz-width bias, and fit a linecenter-offset for each E$^{\prime\prime}$-bin to compensate for pressure-shift bias. These fit parameters reduce the spectral-fit residual, enabling a more accurate normalized linestrength determination from the fit. Appendix \ref{sec:ntfit} defines the lineshape-correction parameters and equations used in the NTfit code \cite{ntfit}.

\begin{figure}
 \centering
 \includegraphics{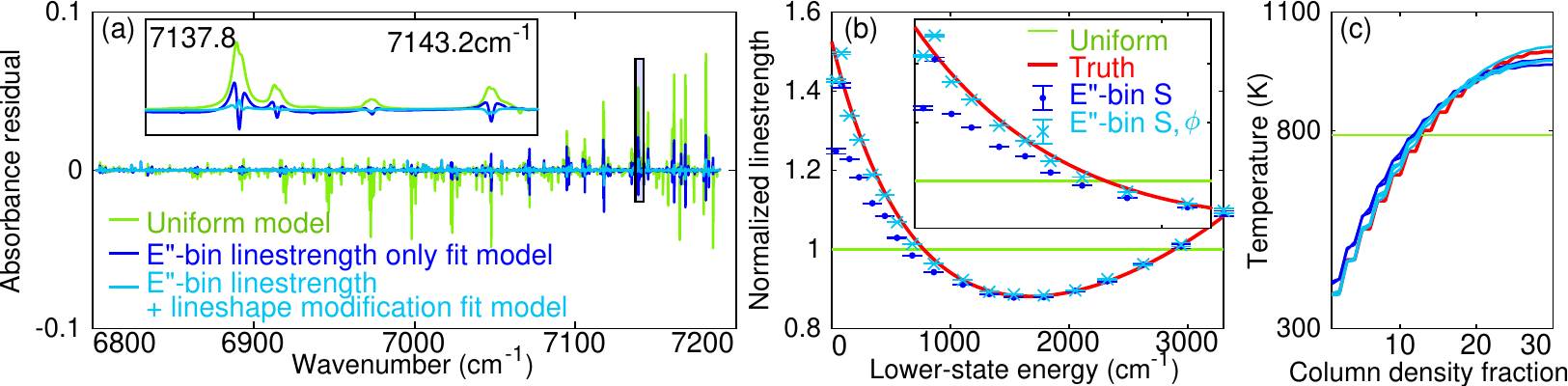}
 \caption{Improvement in  (a) the transmission spectrum fit residual, (b) retrieved normalized linestrengths, and (c) retrieved temperature distribution from fitting a Lorentz linewidths scale factor for each E$^{\prime\prime}$-bin to the Boundary Layer test case from Fig. \ref{fig:noiseless}. Inset to (a) shows residuals at 7140 cm\textsuperscript{-1} of several features with E$^{\prime\prime}$$\approx$ 300 cm\textsuperscript{-1}, where the initial linestrength-only fit has characteristic gull-wing residual of absorption model linewidths that are too narrow. Inset to (b) shows particular improvement in normalized linestrengths for E$^{\prime\prime}$$<$1300 cm\textsuperscript{-1}.}
 \label{fig:lineshape}
\end{figure}

We demonstrate the lineshape refinement to the spectral-fitting algorithm in Fig. \ref{fig:lineshape}.
After the initial spectral fit, the dark blue trace copied from Fig. \ref{fig:noiseless}, we fit a unique pressure and frequency-shift to each E$^{\prime\prime}$-bin to scale the Lorentz-widths and pressure-shifts to produce the light blue traces.
The resulting light blue trace corrects $>$75\% of the normalized linestrength and absorbance residual amplitude error on the low-E$^{\prime\prime}$ lines.
Fitting just the normalized linestrengths first (Table \ref{tab:alg}, step 3) before adding the lineshape modification in a second fit iteration is sometimes more stable than fitting the lineshape right away.

The remaining residual in the light blue trace of Fig. \ref{fig:lineshape} is due to remaining lineshape error, because the nonuniform absorption spectrum is ultimately the superposition of spectra at different temperatures and therefore different linecenters and linewidths.
This sum of Voigt lineshapes produces a complex non-Voigt lineshape, which a single Voigt lineshape fit can never fully capture.
As this effect produces the largest discrepancies at the center of the lineshape, this non-Voigt aggregate lineshape error is the motivation for the hybrid Voigt fit \cite{sanders_2001}.
In contrast to the hybrid Voigt, our E$^{\prime\prime}$-binning method does not correct for this secondary lineshape error, but we see from Fig. \ref{fig:lineshape} that this error does not substantially distort the fit for our test cases.

\section{Proof-of-concept furnace measurement}
\label{sec:expt}

Next, we demonstrate the full temperature inversion algorithm on dual frequency-comb absorption measurements through a laboratory tube furnace.
We recover different temperature distribution shapes with an estimated $\sim$40 Kelvin uncertainty, which are consistent with convection conditions through the system.

\subsection{Tube furnace setup and expected nonuniform temperature profile}
\begin{figure}[h]
 \centering
 \includegraphics[width=6.5in]{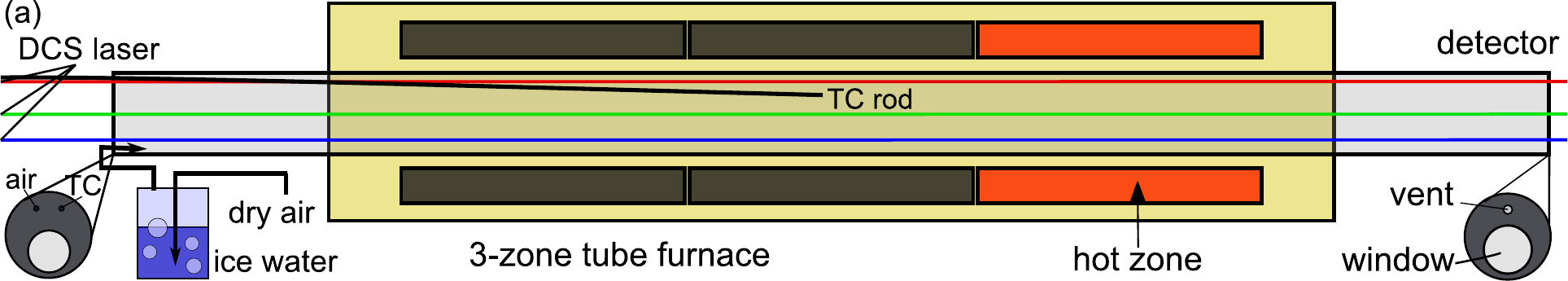}
 \includegraphics{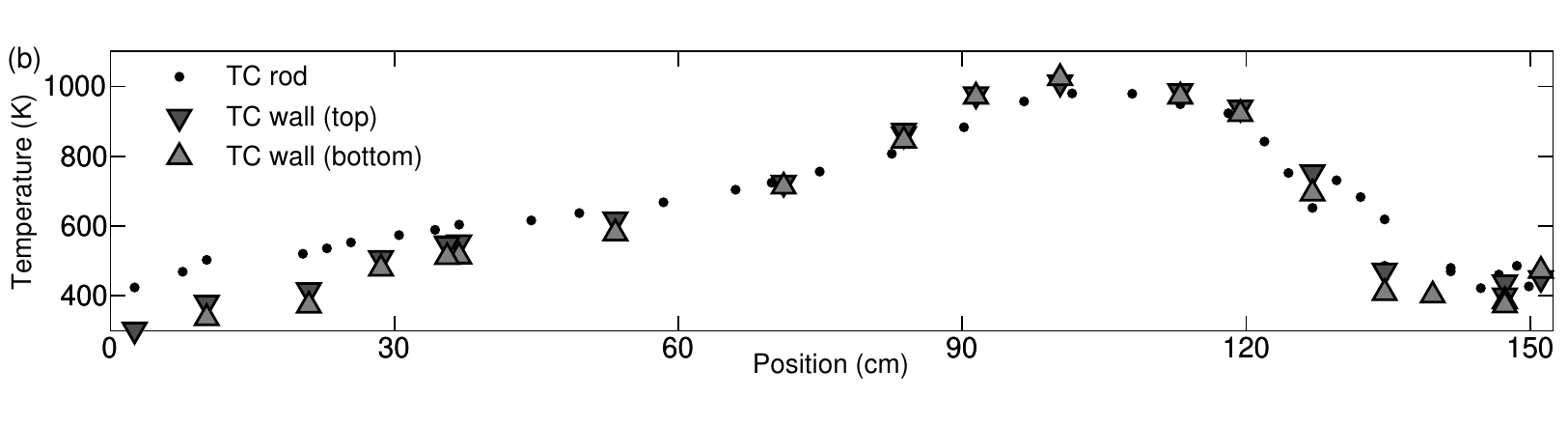}
 \caption{Nonuniform-temperature experiment. (a) Furnace schematic drawn to same length scale as temperature profiles in (b).  Side view of furnace (tan box) with three SiC heaters (rightmost is heated). Bottom corners are orthogonal view insets of end-caps, which are rotated to different window heights for the 3 laser measurement paths (red, green, and blue beams). Steady-state, laminar flow of humid air is supplied from a water bubbler in an ice bath. (b) Thermocouple (``TC") temperature measurements.
The thermocouple rod (dots) is immersed in the gas, and the wall thermocouples (triangles) are attached to the bottom and top outer diameter across the tube.}
 \label{fig:schematic}
\end{figure}

The tube furnace, shown in Fig. \ref{fig:schematic}a, contains three 30 cm long silicon carbide resistive heating zones.
Insulation encloses 107 cm of the 152 cm total tube length, with the remainder protruding from the furnace into the laboratory air.
The 8.6 cm ID fused quartz tube rests in the furnace oven, secured by ceramic supports at 25 cm and 127 cm, with a 1 cm enclosed air gap between the tube and SiC heaters.
To produce a more nonuniform temperature profile, we only power one off-center heater (shown in orange in Fig. \ref{fig:schematic}a), leaving the other two heating zones to be passive insulators.
To maintain steady-state temperature, the tube furnace adjusts the heater duty cycle to maintain a 1000 K reading from a Type N thermocouple floating in the air gap 0.6 cm above the heated tube.
Additional thermocouples monitor the temperature along the outside wall of the quartz tube.
Type K thermocouples provide the ``TC wall” data in Fig. \ref{fig:schematic}b.
In addition to the wall thermocouples, we move a Type K thermocouple rod through the inside of the tube, equilibrating at each location for at least 3 minutes before measurement.
The 2-meter thermocouple (in Omega-clad sleeve with exposed 1.1 mm bead) slid into the tube through a flush 0.6 cm Swagelok connector in the end cap.

Thermocouple rod measurements inside the tube trend about 100 K warmer than wall measurements on the exposed edges and 50 K cooler than wall measurements in the heated zone.
This mismatch between wall and gas thermocouple measurements exceeds uncertainty due to radiation view factors (see Appendix \ref{sec:appendix_nc}), and is indicative of natural convection cells \cite{jones_thermocouple, leong_convection}.
The thermocouple rod measurements are thus biased to some intermediate temperature between the wall and the gas due to the energy balance of radiation and convection at the thermocouple bead tip \cite{jones_thermocouple}.
This temperature bias makes the thermocouple rod unsuitable for determining the gas temperature profile, so we will instead use a natural convection model to evaluate the laser measurement.
In Section \ref{sec:nc}, we will use spectroscopy measurements to derive the boundary conditions of a natural convection model.
We can then compare those wall-temperature boundary conditions against the wall thermocouple measurements, which should not be biased by convection.

To produce a homogeneous, steady-state H\textsubscript{2}O concentration for the dual-comb laser measurement, we force a 24 cm\textsuperscript{3}/s flow (average velocity 0.5 cm/s, Reynolds number 30, see Appendix \ref{sec:appendix_nc}) of cold humid air through the tube.
The laser measurements in Section \ref{sec:nc} will show that this flow rate is too low to disrupt the natural convection cells.
We produce the humid air by bubbling air dried with -40C desiccant through distilled water immersed in an ice bath.
The humid air enters the tube furnace through a 0.6 cm diameter inlet tube on one aluminum end-cap, with a matching 0.6 cm outlet tube on the opposite-side end-cap.
All laser and wall-thermocouple measurements were conducted at least 1 hour into forced-air operation.

\subsection{Dual frequency comb measurement and nonuniform temperature inversion}
\label{sec:expt2}
In this section, we illustrate the full 3-step temperature inversion using the bottom-height laser measurement of Fig. \ref{fig:schematic}a.

Step 1 of the three-step temperature inversion (Fig. \ref{fig:3step}a) is measuring a broadband transmission spectrum.
Dual frequency comb spectroscopy is described in several past works \cite{keilmann_dcs, coddington_dcs_2010, potvin_dcs, ideguchi_dcs, muraviev_dcs}.
We use a fiber-based dual frequency comb spectrometer similar to that described in \cite{zolot_dcs, schroeder_hitemp_hitran}, configured in a single optical pass through the quartz tube onto an InGaAs semiconductor photodetector (Thorlabs PDA10CF).
The spectrometer is capable of measuring the absorption on individual pairs of comb teeth with 200 MHz  (0.0066 cm\textsuperscript{-1}) mode spacing from 6800-7190 cm\textsuperscript{-1}.
This 390 cm\textsuperscript{-1} portion of the transmission spectrum corresponds with an H\textsubscript{2}O absorption database validated at high temperatures. 
Specifically, we use an empirical speed-dependent pure-H\textsubscript{2}O database \cite{schroeder_sdvoigt} together with HITRAN2012 air-broadening spectral parameters. We average the dual-comb data for 10 minutes to reduce spectral measurement noise below $10^{-3}$ absorbance [unitless quantity].  

\begin{figure}[h]
 \centering
 \includegraphics[width=3.9in]{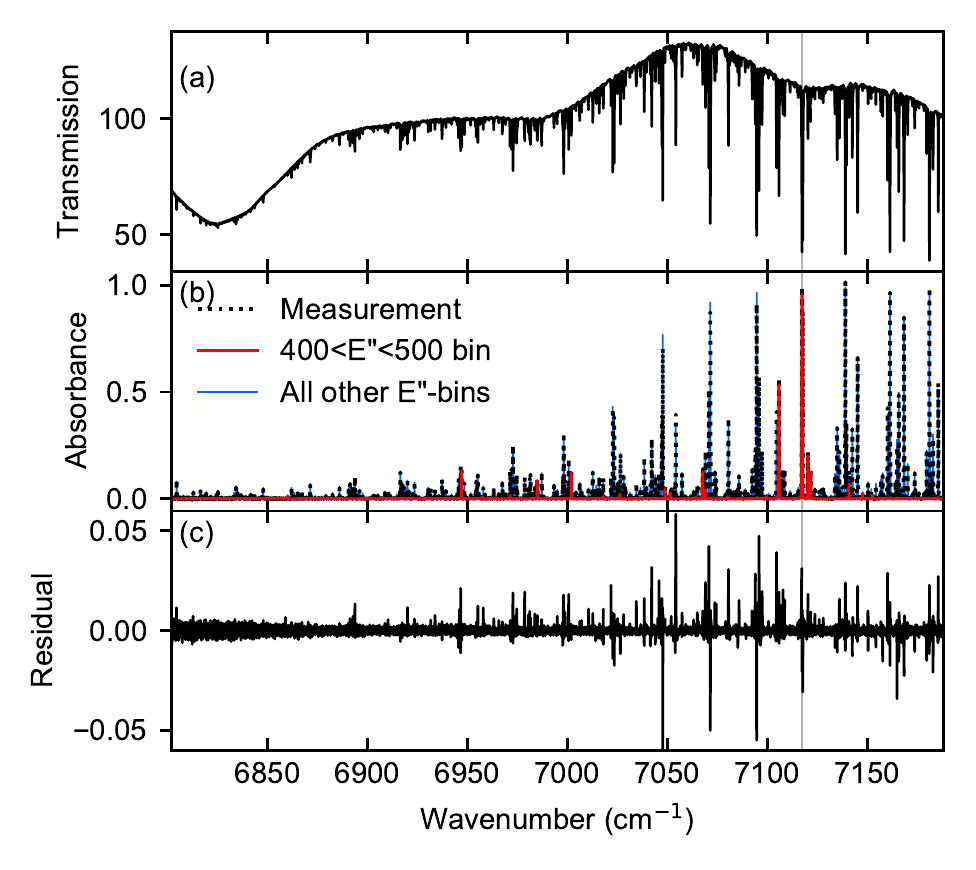}
 \includegraphics[width=2.5in]{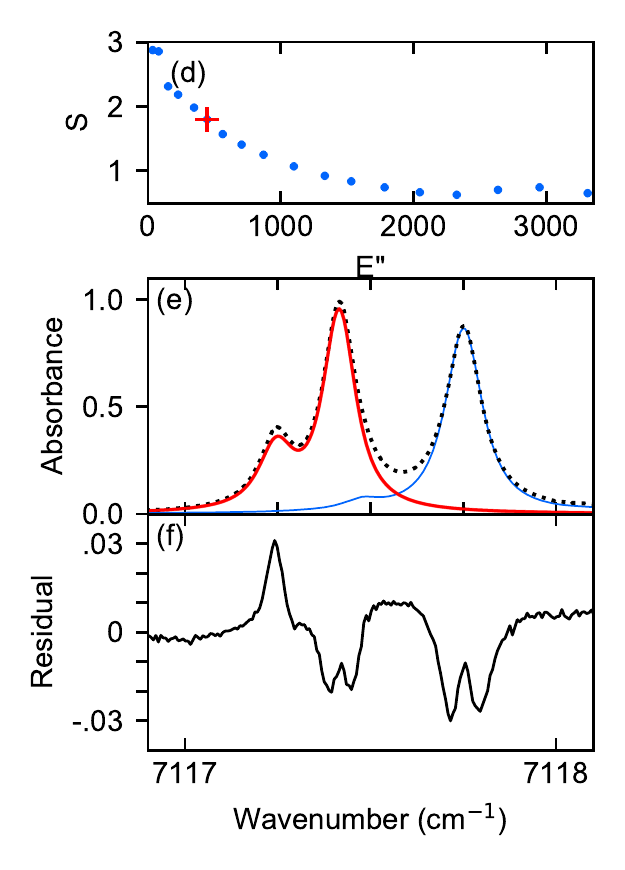}
 \caption{Measured transmission spectrum and fitting results for the bottom height of the nonuniform furnace (blue traces from Fig. \ref{fig:schematic}). Plots from NTfit \cite{ntfit}. (a) Transmission spectrum. (b) Baseline-corrected absorbance spectrum with fit using 18 E$^{\prime\prime}$-bins. The fitted absorption signature for all 400 $<\textrm{E}^{\prime\prime}<$ 500 H\textsubscript{2}O features is shown in red, and summation of other 17 E$^{\prime\prime}$-bins in blue. (c) Baseline-corrected residual (absorbance minus fit). (d) Normalized linestrengths determined from spectral fit in (b), later used to calculate temperature distribution. (e) Inset of the location marked in gray of (b) showing overlapping absorption from features with different E$^{\prime\prime}$.}
 \label{fig:spectrum}
\end{figure}

Step 2 of the three-step temperature inversion (Fig. \ref{fig:3step}b) is extracting normalized linestrengths.
Figure \ref{fig:spectrum} shows this E$^{\prime\prime}$-binning process.
The measured transmission spectrum (Fig. \ref{fig:spectrum}a) is fit (Fig. \ref{fig:spectrum}b) to produce normalized linestrengths (Fig. \ref{fig:spectrum}d).
We use 18 E$^{\prime\prime}$-bins to fit the spectrum, scaling the normalized linestrength, Lorentz-width, and pressure-shift for each bin of H\textsubscript{2}O absorption features to minimize the fit residual (Fig. \ref{fig:spectrum}c).
We use the baseline-free approach described in \cite{time_domain} to fit the transmission spectrum.
To overcome baseline-effects from the 58,000-point spectral fit, we zero-weight the first 100 points of the modified free induction decay signal (m-FID) from the fit, and zero-weight an additional 10 points at the locations in the m-FID signal that correspond to etalons \cite{time_domain}.
This E$^{\prime\prime}$-binning fitting approach corresponds to fitting 57,890 points with 54 free parameters.
This constrained fitting procedure captures most of the absorption signal.
The absorption residuals in Fig. \ref{fig:spectrum}c are of average amplitude 6\% of peak absorbance, although some of this residual is enduring lineshape error rather than normalized linestrength error.
The full residual spectrum has a sum-squared magnitude that is 2.4x the level of the noise floor.
The smooth curve of normalized linestrengths in Fig. \ref{fig:spectrum}d is also indicative of a good quality fit.
This spectral fit also leverages regions (such as Fig. \ref{fig:spectrum}e) of overlapping absorption features with different temperature-dependence to further reduce the statistical fitting uncertainties.

\begin{figure}
 \centering
 \includegraphics[width=6.5in]{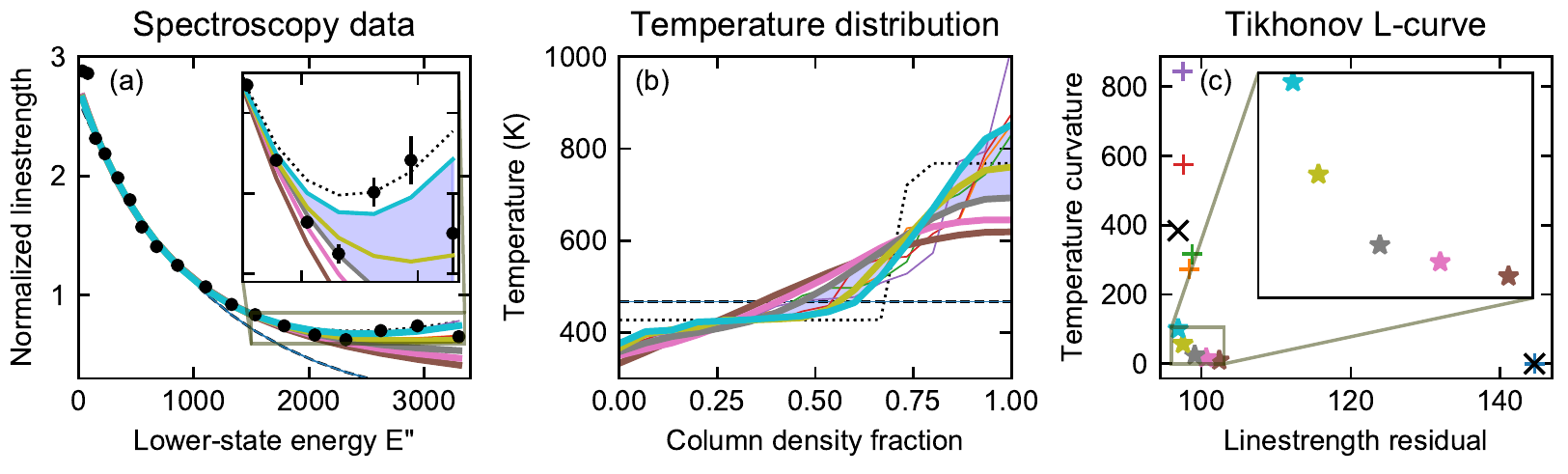}
 \caption{Tikhonov regularization method to determine temperature distribution solution with uncertainty (blue-shaded region of (b)).
(a) Normalized linestrength coefficients from Fig. \ref{fig:spectrum}d shown as black markers.
Each line color represents the corresponding curve-fit for a particular temperature distribution solution in (b), each with a different regularization weight ($\gamma$ from Eq. \ref{eq:tikh}).
(c) L-curve plot is used to determine which temperature distribution solutions to accept.
X-axis values are weighted curve-fit residual from (a), and y-axis values are temperature-distribution-integrated curvatures of solutions in (b).
Select temperature distributions from bottom-left corner of L-curve, representing solutions with smallest linestrength residual (x-axis) as well as small temperature distribution curvature (y-axis).
The parabolic temperature distribution (brown) fits the normalized linestrength with 30\% lower residual than the uniform-temperature fit (black dashes in (a) and (b), x-marker in bottom-right of (c)), and the flat-bottom temperature distributions (yellow-green and teal) are a further 5\% reduction in linestrength residual.
Estimated total temperature distribution uncertainty (blue-shade region of (b)) is the scatter of the 3 flat-bottom temperature distribution solutions within 30\% Euclidean distance from the bottom-left corner of the L-curve of (c).}
 \label{fig:tikh}
\end{figure}

The final step of the temperature inversion (Fig. \ref{fig:3step}c) is solving for the temperature distribution from the normalized linestrengths, using the Tikhonov regularization algorithm in the companion paper \cite{malarich_tx1}.
Briefly, the 2\textsuperscript{nd}-order Tikhonov regularization algorithm determines the temperature distribution which minimizes Eq. \ref{eq:tikh}.
It fits temperature distributions at several values of regularization weight ($\gamma$), selecting the temperature distribution solutions which fit the normalized linestrength data points and have moderate temperature distribution curvature.
\begin{equation}
\label{eq:tikh}
\min \left\{ \lVert \textrm{linestrength residual} \rVert + \gamma \lVert \textrm{temperature curvature} \rVert \right\}
\end{equation}

Figure \ref{fig:tikh} shows the temperature distribution procedure for the bottom height laser measurement.
Each trace (color-coded across the three subplots) indicates a temperature distribution fit at a particular regularization weight $\gamma$, spanning three orders of magnitude.
The x-marks on Figure \ref{fig:tikh}c indicate the high-residual, zero-curvature uniform-temperature fit, and the minimum-residual, high-curvature least-squares fit.
Tikhonov regularization results fall along a hyperbola of the L-curve.
The corner of the L-curve (Fig. \ref{fig:tikh}c inset) includes 5 retrievals whose weighted normalized linestrength residuals vary by 5.3\%.
The temperature distribution plot shows substantial differences for these 5 solutions, ranging from a parabolic distribution with T\textsubscript{max} = 620 K to increasingly negative-skew flat-bottom distributions with T\textsubscript{max} up to 820 K.
The normalized linestrength plot Fig. \ref{fig:tikh}a indicates that this difference in temperature distributions is not solely due to the ill-posedness of the fit, as the curve-fits exhibit considerable difference for E$^{\prime\prime}$$>$2000 cm\textsuperscript{-1}.
The two parabolic distribution fits fall underneath the uncertainty bars of all the E$^{\prime\prime}$-bins in the inset, whereas the three flat-bottom distributions encompass the scatter between the E$^{\prime\prime}$-bins of the inset.
Therefore, we estimate the temperature distribution uncertainty to be the spread of the three flat-bottom solutions, which are the 3 solutions within 3\% linestrength residual in the corner of the L-curve.
This spread in temperature distribution fits is driven by the higher scatter and statistical uncertainty of the E$^{\prime\prime}$$>$2000 cm\textsuperscript{-1} bins in the fit.

\subsection{Temperature inversion algorithm discussion}
\label{sec:discussion}
\begin{figure}
  \centering
  \includegraphics[width=3in]{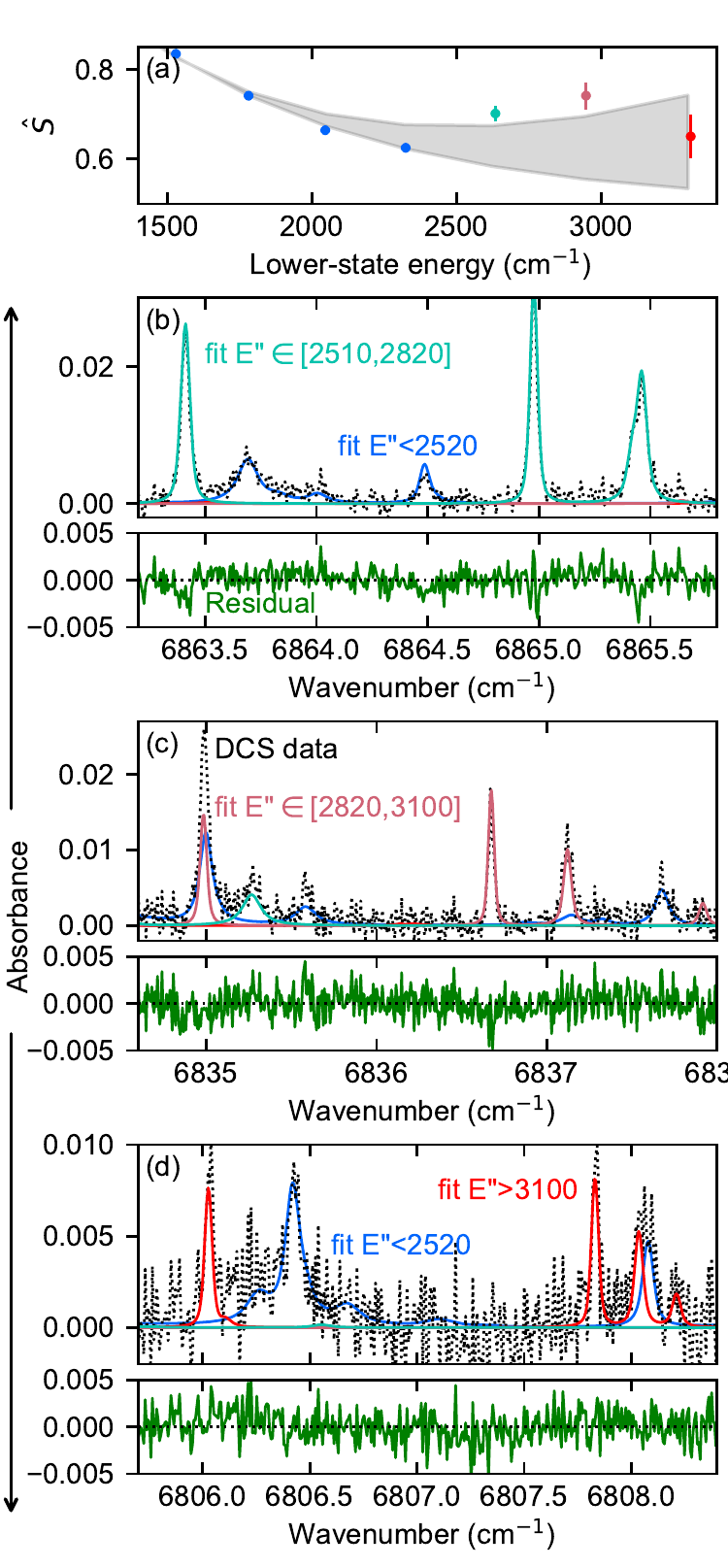}
  \caption{E"-binning fit captures overlapping absorption features with different temperature-dependence. (a) Normalized linestrength inset from Fig. \ref{fig:tikh}a. Markers indicate normalized linestrength coefficients extracted from spectrum with statistical fit uncertainty bars, and shaded region is scatter in curve-fits of possible temperature distributions. (b-d) Insets of spectral fit from Fig. \ref{fig:spectrum} showing the strongest features from the three highest E$^{\prime\prime}$-bins. Fit residuals shown in green beneath each spectrum inset. High-E$^{\prime\prime}$ features at (c) 6835.0 cm\textsuperscript{-1} and (d) 6808.1 cm\textsuperscript{-1} show severe overlap with low-E$^{\prime\prime}$ features (blue), but E$^{\prime\prime}$ fit residual is at noise floor in those regions.}
  \label{fig:precision}
\end{figure}

The bottom-height laser measurement has a larger temperature distribution scatter than the theoretical measurements of the companion paper, due to the low precision of the E$^{\prime\prime}$$>$2000 cm\textsuperscript{-1} normalized linestrength measurements.
This ambiguity in the temperature distribution curve-fit underscores the importance of a laser measurement and spectral fitting technique that maximizes precision.
The high-E$^{\prime\prime}$ absorption features are individually weak for a low-temperature measurement, but the E$^{\prime\prime}$-binning technique fits all of the features.
Fig. \ref{fig:precision} shows spectral fit insets for the three highest E$^{\prime\prime}$-bins. Each inset shows three of the strongest absorption features for each E$^{\prime\prime}$-bin.
Several of the absorption features are overlapping with features from other E$^{\prime\prime}$-bins, having different temperature-dependence.
By fitting an aggregate normalized linestrength rather than individual integrated areas, the E$^{\prime\prime}$-binning technique can easily incorporate those overlapping features to improve the overall fit precision.

We repeated the full procedure with different E$^{\prime\prime}$-bin selections.
Using 43 E$^{\prime\prime}$-bins, we recovered the same flat-bottom temperature distribution result within the estimated uncertainties shown in Fig. \ref{fig:tikh}b.
This fit required more computational time and produced normalized linestrengths with larger scatter and larger statistical uncertainty bars (larger $\sigma_{\hat{S}}$), but did not change the final temperature distribution result.
We also performed fits with 8 and 10 E$^{\prime\prime}$-bins, finding that these fits had a 20\% higher spectral residual 2-norm, and produced the lower-curvature parabolic temperature distribution. These fits are labeled in the supplementary data in the subdirectory ``ebin\_selection” \cite{data_partII}. 

We also measured dual-comb spectroscopy data at a height 10\% below the top of the tube (red traces in Fig. \ref{fig:results}).
The Tikhonov solutions produced from this measurement contained much less scatter than in Fig. \ref{fig:tikh}b--the measurements near the L-curve corner were all flattened-top parabolas, and only varied by 1\% linestrength residual.
This lower temperature distribution scatter for the top height measurement was due to higher SNR for the E$^{\prime\prime}$$>$2000 cm\textsuperscript{-1} bins, as those absorption features were larger for this higher-temperature measurement.

\subsection{Evaluation against natural convection model}
\label{sec:nc}
\begin{figure}
  \centering
  \includegraphics[width=6.5in]{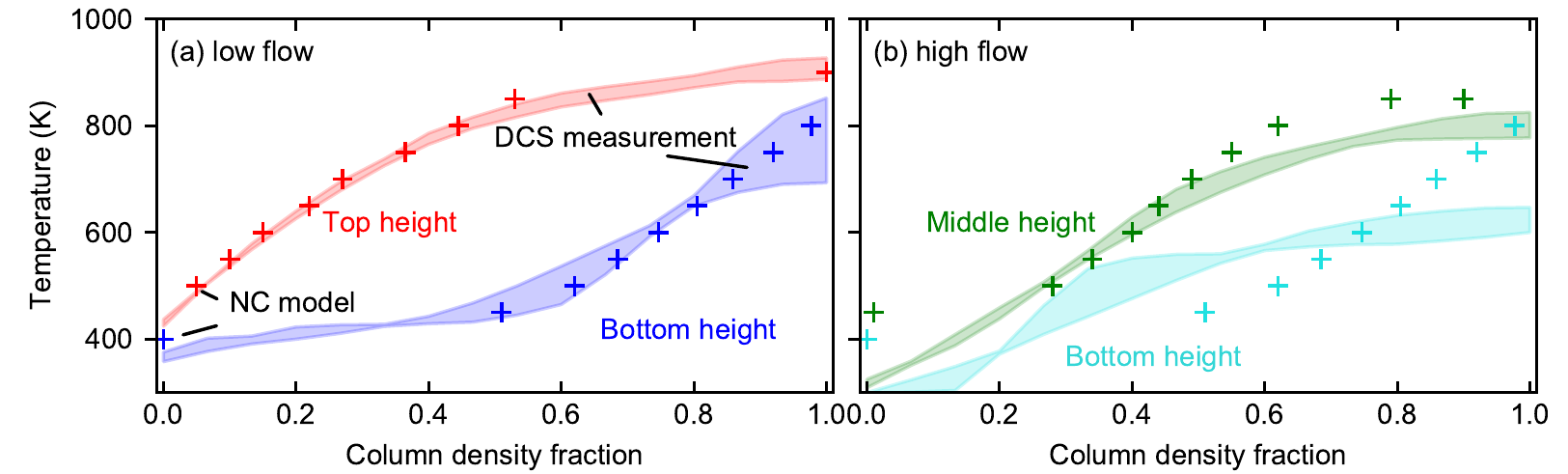}
  \caption{Using DCS temperature distribution measurements (shaded regions) at different heights through the tube furnace to calibrate the boundary conditions of a 3D natural convection model. (a) DCS temperature distributions for cold-air flow rate 24 cm\textsuperscript{3}/s. Red measurement is at height y/D=0.9, and blue measurement is at y/D=0.2. Shaded region indicates the scatter of temperature distribution Tikhonov solutions near the corner of the L-curve (see Fig. \ref{fig:tikh}). Markers indicate natural convection contours from Fig. 5 of \cite{leong_convection} for Ra=20,000 and wall-temperature boundary conditions $T_H$ = 900 K, $T_C$ = 400 K, 50\% heated fraction.
(b) Measurements at higher 70 cm\textsuperscript{3}/s cold-air flow rate do not match temperature distribution predictions from same natural convection model (+ markers). Green measurement is at height y/D=0.55, and cyan measurement is at y/D=0.2.}
  \label{fig:results}
\end{figure}

Figure \ref{fig:results} superimposes the temperature distribution results at two laser path heights: 10\% below the top wall of the tube (y/D=0.9) (red) and 20\% above the bottom (blue).
The top measurement is a 240 K higher mean temperature than the bottom measurement, and a different temperature distribution shape.
Specifically, the top measurement is a negative-skewed flat-top distribution, whereas the bottom measurement is a positive-skewed flat-bottom distribution.
This indicates that the technique applied to the DCS data can determine not just the degree of temperature nonuniformity (or temperature variance), but a more complex temperature distribution shape.

To evaluate the spectroscopy temperature distributions, we use a published natural convection model from the analogous system of a horizontal tube with a hot and cold side \cite{leong_convection}.
We assume an adiabatic boundary condition at x=107 cm, and apply the same model solution on both sides, just over different horizontal length scales.
The tube has a Rayleigh number of $\sim$200,000 (see Appendix \ref{sec:appendix_nc}), and the model predicts that the natural convection cells fill the domain at Rayleigh $>$ 20,000.
Therefore, we produce the ``model” temperature profiles in Fig. \ref{fig:schematic}b from the Ra = 20,000 contour plots in the paper \cite{leong_convection}.
The temperature field is calculated from three boundary conditions: the heated-wall temperature $T_H$, the cooled-wall temperature $T_C$, and the heated-wall fraction $b$.

We adjust these three boundary conditions in order to fit the natural convection model to the DCS temperature distribution measurements.
The Fig. \ref{fig:results} cross marks show the predicted temperature distributions using the best-fit natural convection model boundary conditions: $T_H$ = 900 K, $T_C$ = 400 K, $b$ = 0.5.
In Fig. \ref{fig:results}a, the fitted natural convection model distributions are in good agreement with the DCS temperature distributions.
The natural convection model matches the temperature range of both DCS measurements, as well as the extent of the flat-top and flat-bottom regions.
This shows the promise of using sparse nonuniform-temperature measurements to constrain boundary conditions in gas systems.

\begin{figure}
  \centering
  \includegraphics[width=6.5in]{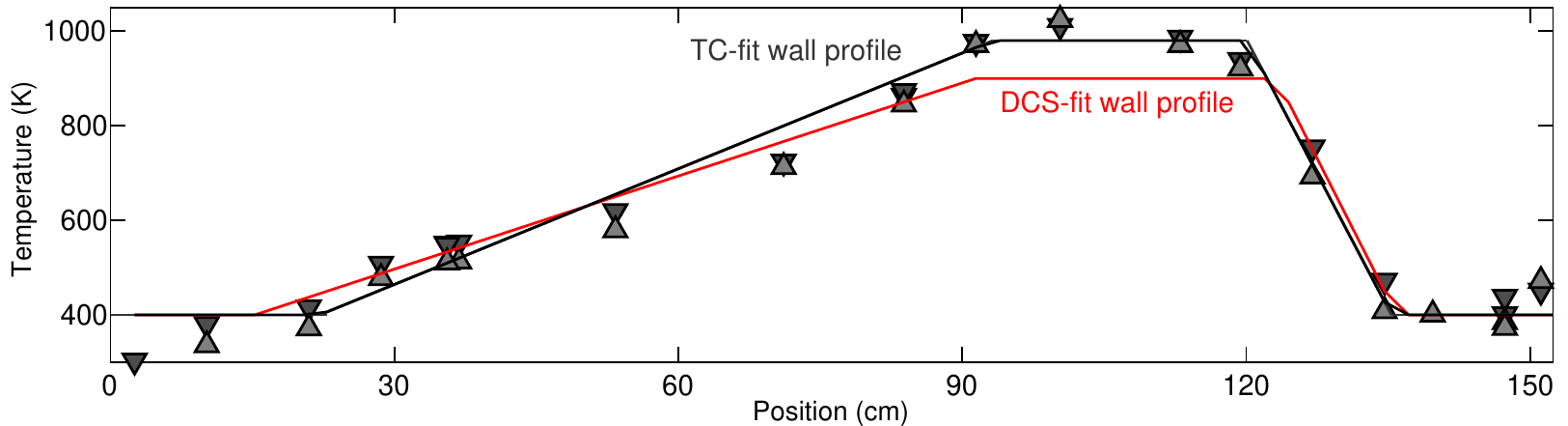}
  \caption{Wall temperature profiles (lines) used for 3D natural convection model from \cite{leong_convection} compared against wall thermocouple measurements (triangles). Grey line is the best-fit boundary conditions to the wall thermocouple data, using $T_C$ = 400 K, $T_H$ = 980 K, and b = 0.46. Red line is the best-fit boundary conditions to match the natural convection model temperature distributions to low-flow DCS data (from Fig. \ref{fig:results}a, using $T_C$ = 400 K, $T_H$ = 900 K, and b = 0.5.}
  \label{fig:walltc}
\end{figure}

To test the plausibility of the laser temperature distributions, we compare the spectroscopy-tuned boundary conditions against the wall thermocouple data.
First we note the naive boundary conditions: the 152 cm tube includes a 30 cm section exposed to a heater at 1000 K, and two 23 cm sections exposed to 298 K laboratory air.
If we attribute half of the remaining 83 cm tube section to the heated wall and half to the cooler wall, this corresponds to a hot fraction $b$ = 0.47, which is close to the measured $b$ = 0.5.
In Fig. \ref{fig:walltc}, the wall thermocouples varied between 400 K and 1000 K. A best-fit wall thermocouple model has $T_C$ = 400 K, $T_H$ = 1000 K, and $b$ = 0.46.

The radiative correction of all of the thermocouples could be particularly complicated around $T_{max}$.
The 0.33 cm-thick quartz tube is 45\% transparent to a 1000 K blackbody spectrum. So all of the thermocouples are in radiative balance with not only the quartz tube wall, but also the silicon carbide heaters outside the tube. The quartz wall ID is cooled by the substantial natural convection cells in the tube flow, so it could be substantially colder than the silicon carbide heaters that are entirely encased in insulation. 
The natural convection model predicts a cooling rate of about 6000 W/m\textsuperscript{2}, which would correspond to a 20 K temperature gradient across the 0.6 cm quartz tube wall, and a 40 K difference from radiation between the semi-transparent quartz tube and the silicon carbide heaters.
The convection model boundary conditions are for the inner diameter, whereas the thermocouples are loosely tied to the outer quartz wall.
Thus vertical gradients justify an inner-diameter boundary condition of 940 K, which is 5\% higher than the DCS prediction.
The other boundary conditions, $T_C$ and $b$, are within wall thermocouple measurement precision.

The natural convection model does not take into account H\textsubscript{2}O molefraction nonuniformity. The DCS temperature distribution has an x-axis ``column density fraction” that describes the pressure-molefraction-pathlength fraction of H\textsubscript{2}O molecules below a temperature $T$. Any path-nonuniformity in molefraction should not affect the temperature range for $T_C$ and $T_H$, but it could alter the shape of the temperature distribution. However, the fact that both DCS and wall-thermocouple measurements support a hot-fraction $b$=0.5 suggests that the water molefraction is homogeneous in this system, because wall thermocouples should not be sensitive to concentration gradients.

We also recorded two DCS measurements at a 3x-higher forced-air flowrate (70 cm\textsuperscript{3}/s, giving an average forced-air speed of 1.5 cm/s through the tube).
Fig. \ref{fig:results}b shows the resulting temperature distributions at two heights, including one shared height with the low-flow rates in Fig. \ref{fig:results}a.
These measurements do not match the same natural convection model as the low-flow measurements in Fig. \ref{fig:results}.
In particular, both DCS measurements in Fig. \ref{fig:results}b predict a minimum temperature of 300 K, colder than the 370 K minimum temperature of the low-flow bottom height in Fig. \ref{fig:results}a.
These results are an indication that the cold forced-air flow rate is high enough to alter the natural convection cell upstream of the heated region.
In fact, wall-thermocouple measurements upstream of the heater are 20 K colder for this higher-flow condition.

\section{Influence of bandwidth}
\label{sec:bandwidth}
The companion paper found that there are diminishing returns to the accuracy of the retrieved temperature distribution for theoretical measurements using more than $\sim$15 well-selected absorption features \textit{in the absence of error}.
Table \ref{tab:errors} includes several sources of error that are introduced by real-world experiments and data analysis, and were not considered in the companion paper. 
Three of these sources of error decrease as the algorithm incorporates more absorption lines (through a larger measurement bandwidth): measurement noise, spectral parameter error, and error due to nonuniform laser intensity baseline and overlapping absorption features.
In this section, we will argue from first principles and Monte Carlo simulation how extra bandwidth incorporating more than 15 absorption features improves the temperature distribution accuracy in real-world absorption measurements.

\subsection{Theory of bandwidth improvements}
The three experimental error sources (noise, out-of-system background absorption, and simulation/experiment wavelength matching) degrade the E$^{\prime\prime}$-binning fit in the same way they would affect a multi-species broadband fit. In a multi-species measurement, these error sources alter the retrieved molefractions; in this nonuniform-temperature fit, these errors alter the retrieved normalized linestrengths at each E$^{\prime\prime}$-bin.
Lower uncertainties in normalized linestrengths should propagate to higher temperature distribution accuracy, although this propagation is more difficult to quantify due to the complexity of nonlinear regression with Tikhonov regularization. 

The detection limit \cite{adler_mid-ir} determines the statistical (Jacobian-estimated) uncertainty of the normalized linestrength fit from each individual E$^{\prime\prime}$-bin. Spectral noise introduces uncertainty into the normalized linestrengths via Eq. \ref{eq:detectionlimit}.
\begin{equation}
\label{eq:detectionlimit}
\sigma_{\hat{S}} = \frac {\sigma_\alpha} {\sqrt{\sum_{i=bl}^N \alpha_{\hat{S}}^2(t_i) }}
\end{equation}

Here $\sigma$ is noise, $\alpha$ is absorbance, and $\alpha(t_i)$ is the time-domain absorption of all the features in one E$^{\prime\prime}$-bin, summed over the portion of the m-FID spectrum that is uninfluenced by laser intensity baseline, $bl$.
As the spectral bandwidth increases, the additional absorption features increase the summation term in the denominator of Eq. \ref{eq:detectionlimit} and thus reduce the uncertainty of the normalized linestrength $\sigma_{\hat{S}}$.
This relationship is derived for single-molecule spectral fits -- a multivariate fit for multiple E$^{\prime\prime}$-bins would have some correlation term to increase the fit uncertainty when the spectrum contains overlapping absorption features from different E$^{\prime\prime}$-bins.
More bandwidth should reduce this correlation effect, as the relative sizes of overlapping absorption features will scale with other isolated absorption features elsewhere in the spectrum. 

Measuring additional features should also reduce the magnitude of normalized linestrength error induced by spectral parameter error.
Each absorption feature produces some normalized linestrength according to a fit of the integrated area divided by the reference linestrength.
Random errors in the lineshape parameters and $S_{296}$ reference linestrength produce error in the normalized linestrength fit, which averages down as more features are incorporated into each E$^{\prime\prime}$-bin. 
For instance, H\textsubscript{2}O spectral parameters in the HITRAN database \cite{hitran2016} have a source file that states for certain self-widths ``polynomial fit of the values from \cite{antony_2007},'' which indicate the database could be ignoring normally-distributed scatter in the parameters.
In reality, spectral parameters could also have some systematic error component shared across all absorption features, for instance due to error in the gas cell conditions for empirical databases.
Extra bandwidth will not reduce this systematic error source.

\subsection{Simulation of bandwidth-dependence}
The normalized linestrength precision only matters insofar as it affects the retrieved temperature distribution accuracy.
Qualitatively, more spectral bandwidth should produce a more precise normalized linestrength curve (as described above), which in turn should reduce the uncertainty of the nonuniform temperature distribution result.
However, this link to the temperature distribution is difficult to determine from first principles, particularly when the temperature inversion step includes regularization.
Therefore, we instead present a Monte-Carlo simulation of how bandwidth helps a realistic measurement.
The Monte-Carlo framework shows how statistical uncertainty in both measurement and spectral parameters alter the temperature result from a synthetic truth, and how the same statistical uncertainty degrades the temperature result less for larger spectral bandwidths.

\begin{figure}[h]
 \begin{subfigure}[tl]{6cm}
\begin{small} 
\vskip 0pt
 \begin{tabular}[t]{cx{12mm}c cx{7mm}c cx{6mm}c}
  \toprule
 \makecell[tc]{Bandwidth \\ (span) (cm\textsuperscript{-1})} & \# \ lines & \# \ E$^{\prime\prime}$ \\
 \midrule
 6916-6923 (  7) & 20 & 12 \\
6916-6926 ( 10) & 28 & 15 \\
6916-6936 ( 19) & 41 & 18 \\
6900-6940 ( 40) & 74 & 18 \\
6900-6950 ( 50) & 98 & 18 \\
6802-6950 (148) & 241 & 18 \\
6802-7150 (348) & 697 & 18 \\
 \bottomrule
 \end{tabular}
\end{small}
\end{subfigure}
\begin{subfigure}[tr]{5cm}
 \vskip 0pt
 \includegraphics{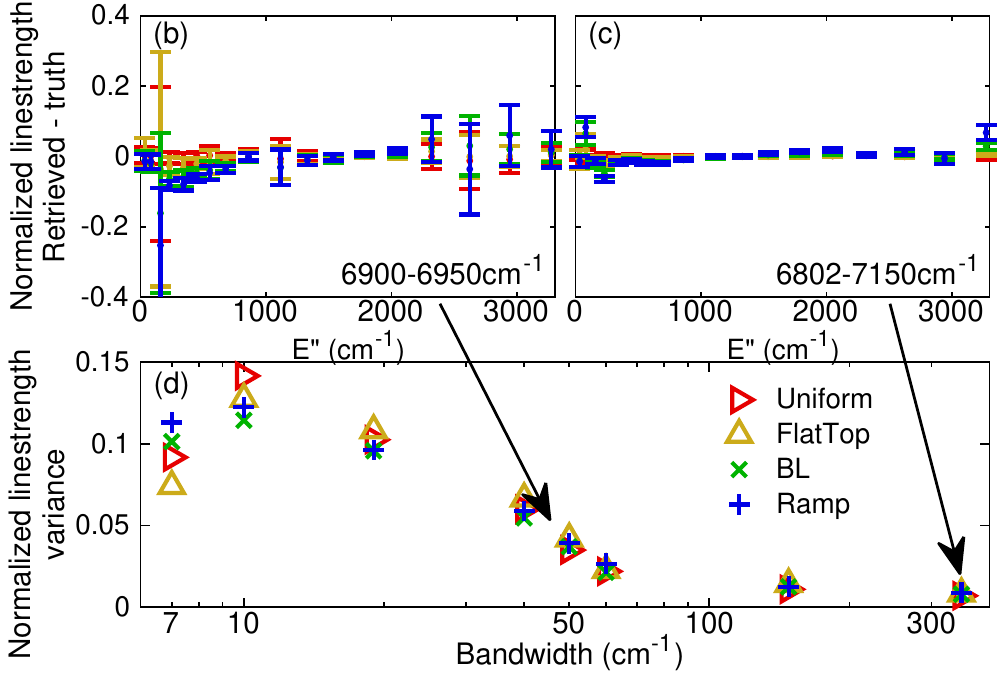}
\end{subfigure}
 \caption{Bandwidth conditions and normalized linestrength measurements from Monte-Carlo study. (left) List of test bandwidths, which start at a strong 2-absorption-feature region at 6916 cm\textsuperscript{-1} and increase. (right) Deviation and scatter in recovered normalized linestrength over N=50 trials for different bandwidths. All bandwidths have negligible normalized linestrength bias, but larger bandwidths have less scatter.}
 \label{fig:variance}
\end{figure}

In this Monte-Carlo study, we simulate H\textsubscript{2}O transmission spectra for the four test-case temperature distributions at eight different spectral bandwidths.
We add noise and spectral parameter error to the synthetic spectrum, and use NTFit \cite{ntfit} to retrieve the normalized linestrengths and then the temperature distribution.
As a Monte-Carlo study, we repeat this process 50 times for each bandwidth, each with a different measurement noise and spectral parameter error.
The spectral parameters were given a one-standard-deviation uncertainty of 1\%, to match the statistical fitting uncertainty in Schroeder et al. \cite{schroeder_sdvoigt}.
The reference linestrength, linecenter, foreign broadening width, lineshift, and temperature-dependence exponents were given a random different value for each of the 50 synthetic measurements.
A different 0.05\% transmission noise was applied to each synthetic measurement, along with a 0.2\%-transmission baseline etalon with random phase, to match the measurement conditions in Section \ref{sec:expt}.
The bandwidths selected for this study match the conditions in the companion paper, where the initial measurement of the two-line pair at 6917 cm\textsuperscript{-1} was broadened in 8 steps to incorporate the 6802-7150 cm\textsuperscript{-1} region used in the empirical spectral parameter study of \cite{schroeder_sdvoigt}.
The selected bandwidths are described in the table in Fig. \ref{fig:variance}.
The ``number of lines'' column approximates the number of absorption features above the noise floor, defined as a feature with line intensity $S(T)$ exceeding a linestrength threshold of 2x10\textsuperscript{-23} cm/molecule at any temperature in the fit range of 300-1300 K.
This threshold corresponded to an average of approximately two absorption features for each wavenumber of bandwidth. 

As the bandwidth increased from 4 to 20 cm\textsuperscript{-1}, strong quantum transitions with different E$^{\prime\prime}$ were incorporated into the fit.
The four subsequent bandwidths did not add additional E$^{\prime\prime}$-bins, but instead added additional lines (redundancy) to each E$^{\prime\prime}$-bin.
These extra absorption lines at larger bandwidths did reduce the scatter of the retrieved normalized linestrengths, as shown in Fig. \ref{fig:variance}d.
The stronger high-temperature absorption features in 6800-6900 cm\textsuperscript{-1} particularly helped reduce the retrieved linestrength scatter for E$^{\prime\prime}$$>$2200 cm\textsuperscript{-1}.
As the ambiguity in E$^{\prime\prime}$$>$1500 cm\textsuperscript{-1} drove the temperature distribution uncertainty in the bottom path (Fig. \ref{fig:results}d), we expect this reduction in retrieved linestrength scatter to improve the temperature distribution inversion.

\begin{figure}[h]
 \centering
 \includegraphics{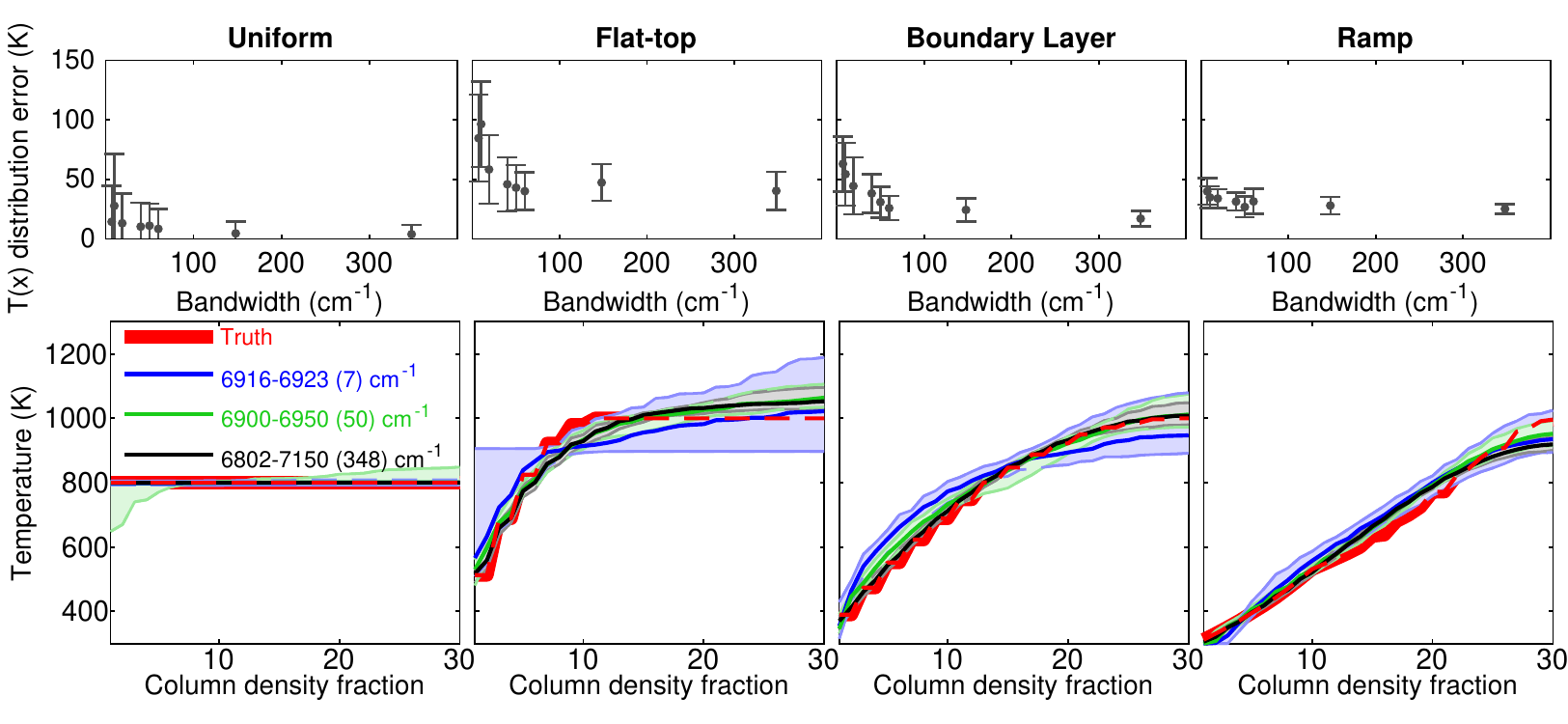}
 \caption{Temperature inversions from synthetic noisy transmission spectra in Fig. \ref{fig:noiseless}. (top) Mean distribution error (N=50 trials) decays as more absorption lines are incorporated into spectral fit. (bottom) Mean temperature distributions for three of the bandwidths in the study. Shaded error bars are 15-85\textsuperscript{th} percentile temperatures recovered for each length bin.}
 \label{fig:mc}
\end{figure}

Fig. \ref{fig:spectrum} shows the temperature distribution statistics from the Monte Carlo study.
The top row shows path-integrated temperature distribution error for each bandwidth. 
A large distribution error indicates that the temperature distributions were consistently far from the true distribution across all length bins.
Large uncertainty bars indicate the temperature retrieval varied for different Monte-Carlo runs.
The bottom row of Fig. \ref{fig:mc} shows the average recovered temperature distribution across all Monte-Carlo runs for three different bandwidths.
The shaded error bars show the 15-85\textsuperscript{th} percentile temperatures across the 50 Monte-Carlo runs.
The smallest bandwidths produced inaccurate temperature distributions with large scatter for different runs.
As bandwidth increased, temperature accuracy converged near the true distribution for three of the four test cases.
The ramp distribution had a similar scatter and distribution shape for all bandwidths.
This stability comes from the Tikhonov regularization condition, which biases the retrieval to a similar profile even when the normalized linestrengths have more scatter.
For the other three profiles, the distribution-averaged mean temperature error shown in the top row of Fig. \ref{fig:mc} show significant improvements up to 50-100 cm\textsuperscript{-1}, with additional bandwidth reducing the scatter across the different inversion trials.
The uniform and boundary-layer cases continue to improve up to the largest 350 cm\textsuperscript{-1} bandwidth. 

Another temperature distribution accuracy metric is how frequently the inversion predicts whether there is nonuniformity.
We force the inversion to return a uniform-temperature result when allowing temperature nonuniformity improves the normalized linestrength fit by less than 7\% \cite{malarich_tx1}.
The uniform test-case has 15\% false-positives, meaning that the inversion incorrectly predicted that the distribution was nonuniform for 15\% of the trials. All bandwidths had a similar false-positive rate, however, the smaller bandwidths predicted more substantial temperature nonuniformity with their false positives. The smallest bandwidths also had larger mean temperature scatter in their true negative retrievals, frequently exceeding 20 Kelvin offset. The flat-top and boundary-layer cases also had some false negatives. The flat-top case incorrectly produced a uniform-temperature retrieval for 20\% of the Monte-Carlo runs at the smallest bandwidths. The bandwidths spanning more than 10 cm\textsuperscript{-1} produced no false negatives in the study, which used a small 0.05\% transmission noise.

Although the additional bandwidth continued to reduce the scatter and accuracy of the normalized linestrengths in Fig. \ref{fig:variance}, those gains did not consistently translate to increases in temperature inversion accuracy. Not all test cases benefited from the largest bandwidths. However, in a real measurement where the true distribution shape is uncertain, the extra bandwidth and lower normalized linestrength scatter can add confidence to the final human interpretation of the temperature inversion.

\section{Conclusions}
We developed a post-processing method to recover the spatial temperature distribution from a single-beam broadband absorption spectroscopy measurement.
A companion paper identified the lower-state-energy-dependent normalized linestrengths as the core aspect of the spectroscopy measurement that provides temperature nonuniformity information.
This paper introduces a constrained spectral fitting method, E$^{\prime\prime}$-binning, to extract these normalized linestrengths from a raw transmission spectrum.
The spectral fitting method is available from \cite{ntfit}.
This full-spectrum fitting method, applied to a transmission spectrum congested by overlapping water vapor absorption features, provides an automated method to leverage the entire laser measurement.
We show that this spectral fitting method combined with the Tikhonov regularization technique (from the companion paper \cite{malarich_tx1}) produces robust temperature distributions in high temperature conditions in both simulated and experimental demonstrations.

We demonstrate the full nonuniform-temperature technique on several dual frequency comb laser measurements through a laboratory tube furnace. Despite absorption model uncertainty and measurement noise, we recover distinct temperature distribution shapes for the different laser measurements. We fit a natural convection model to the temperature distributions by adjusting boundary conditions. 
These spectroscopy-determined boundary conditions match the wall thermocouple measurements to within 5\%, albeit with a larger -80 K bias in the heated region where wall thermocouple measurements are less informative due to wall transmissivity and multi-mode heat transfer. Moreover, we compared temperature distribution measurements at different flow conditions to determine at which forced-air flow rate the pure natural convection model breaks down. These measurements demonstrate the potential of the technique to recover a variety of temperature distributions and to quantify relevant heat transfer phenomena in unknown heterogeneous flows.
The ambiguity in the bottom-height measurement of the tube furnace illustrated the importance for the spectral fitting method to extract accurate, precise normalized linestrengths across a wide range of lower-state energy.
Adding spectral bandwidth to measure many more than 15 absorption features improves the normalized linestrength precision and in turn reduces the temperature distribution uncertainty.

In principle, the inversion method described in this two-part paper series should also work for more complex environments, such as combustion and reacting systems.
However, the even higher temperatures and composition nonuniformities in, for example, combustion systems, add challenges to this nonuniform-temperature algorithm.
In particular, these conditions introduce additional uncertainty into the mean-temperature Voigt lineshape approximation, which could be mitigated with iterations to the spectral fitting method.
Another challenge is measuring systems that contain nonuniformity in concentration and potentially pressure in addition to temperature.
In some instances, the normalized linestrength-to-temperature distribution step may need to accommodate a variable concentration profile.
This higher-parameter inversion can be stabilized by incorporating expectations on the interrelated structure of the composition and temperature profiles in combustion systems.
In practice, this extra output requirement is likely to increase the uncertainty on the retrieved temperature profile.
There are opportunities for future algorithm developments that improve uncertainty quantification and nonuniform-concentration retrievals.
This technique enhances the utility of broadband lasers for absorption-based measurements in many nonuniform environments.
\\
\bigskip
\\
\textit{Acknowledgments}: This work was funded by the National Science Foundation under grant CBET 1454496 and the Air Force Office of Scientific Research under grant FA9550-17-1-0224. Additionally, we thank the anonymous reviewers for their detailed comments which greatly improved this work.

\bibliography{zotero_library}
\bibliographystyle{elsarticle-num}

\appendix
\renewcommand{\theequation}{\thesection.\arabic{equation}}
\renewcommand{\thefigure}{\thesection.\arabic{figure}} 
\renewcommand{\thetable}{\thesection.\arabic{table}}
\section{Availability of material}
The full temperature inversion algorithm, the $NTfit$ package, is available openly under the BSD-3 license \cite{ntfit}. The figures in this article, as well as the data and plotting scripts used to produce them, are available openly under the CC-BY license \cite{data_partII}.

\setcounter{equation}{0}
\setcounter{figure}{0}
\setcounter{table}{0}
\section{E"-binning with modified Fourier Induction Decay fitting}
\label{sec:mfid}
Section \ref{sec:ebin} introduces E"-binning to fit absorbance spectra. This section describes how to adapt the equations to fit a transmission spectrum containing a laser intensity baseline. This transmission spectrum fit technique is based on m-FID fitting \cite{time_domain}, and preserves the linearity of the E"-bin fit.

M-FID fitting is a technique to avoid the influence of the native laser intensity baseline on the spectrum fit, using Eq. \ref{eq:mfid}.
\begin{equation}
\label{eq:mfid}
W \times \mathcal{F}^{-1} [-\textrm{ln}(I\textsubscript{meas})] = W \times (\mathcal{F}^{-1}(\alpha\textsubscript{meas}) + \mathcal{F}^{-1}[-\textrm{ln}(I_0)]) \sim W \mathcal{F}^{-1}(\alpha\textsubscript{meas})
\end{equation}
where $I\textsubscript{meas}$ is the raw spectroscopy measurement, and $\alpha\textsubscript{meas}$ is the same gas absorbance measurement shown in Eq. \ref{eq:snorm}.
With the appropriate weighting function $W$, one can remove the laser intensity spectrum $I_0$ from the measurement.

The normalized linestrength fit shown in Eq. \ref{eq:snorm_fit} is written in terms of the measured absorbance $\alpha\textsubscript{meas}$, but one can take the Fourier transform of Eq. \ref{eq:snorm_fit} then substitute the transmission spectrum using Eq. \ref{eq:mfid}.
Due to the linearity of the Fourier Transform, we can rewrite this minimization as follows:

\begin{equation}
\label{eq:mfid_linear}
\{ \hat{S}_i^{est}  \} = \textrm{arg} \min_{\hat{S}_i} \sum_{t}^M \left( W(t) \times \left\{ \mathcal{F}^{-1} [-\textrm{ln}(I\textsubscript{meas})] - \sum_i^N \hat{S}_i \mathcal{F}^{-1}[ \alpha_{0,i}]   \right\}  \right)^2
\end{equation}

The absorption signatures for each E"-bin, $\mathcal{F}^{-1}[ \alpha_{0,i}] $, are precomputed from a line list containing all of the H\textsubscript{2}O absorption features in some range of E".
Thus the second term of Eq. \ref{eq:mfid_linear} is an M$\times$N matrix, and $\hat{S}_i$ are the unknown linear fit coefficients.
The weighting function $W(t)$ is a high-pass filter, with a value of zero up to some threshold time, and otherwise a value of unity save for individual etalons.
Thus Eq. \ref{eq:mfid_linear} can be rewritten as min($A_w x - b_w$), and then solved by linear regression $x_{est} = (A_w^T A_w)^{-1} A_w^T b_w$. Here $A_w$ is the $M\times N$ matrix $W \mathcal{F}^{-1} [\alpha_{0,i}]$ and $b_w$ is the $M\times 1$ vector $W \mathcal{F}^{-1}[-ln(I\textsubscript{meas})]$.

\begin{figure}[h]
\centering
\includegraphics[width=3in]{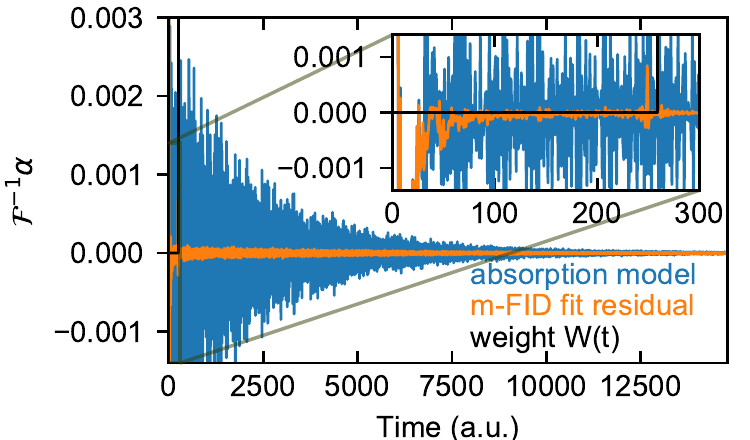}
\caption{Modified Fourier Induction Decay fit for spectrum shown in Fig. \ref{fig:spectrum}a.
Blue trace shows aggregate fit model from all E$^{\prime\prime}$ bins, black trace shows weight vector $W(t)$, and orange trace shows fit residual from Eq. \ref{eq:mfid_linear}.
Inset shows first 300 points, where residual contains laser intensity baseline that is unweighted for the m-FID fit.}
\label{fig:mfid}
\end{figure}

Fig. \ref{fig:mfid} shows the time-domain fit for the bottom-height DCS measurement from Fig. \ref{fig:spectrum}. The absorption signature (blue) is distributed in time, whereas the laser intensity baseline is localized at early times. By applying a weight function $W(t)$ (black trace) set to zero for the first 250 points, the m-FID fit can fit most of the absorption signal free from the influence of the baseline and etalons.

This m-FID approach does not directly fit the absorbance spectrum; however, an approximate absorbance spectrum can also be recovered from this fit.
This absorbance spectrum, $\alpha\textsubscript{meas}$ shown in Fig. \ref{fig:spectrum}b, is recovered following the baseline normalization procedure in Chapter 8 of \cite{malarich_dissertation}. 

\begin{equation}
\label{eq:i0}
\alpha(\nu) \approx -\textrm{ln}(I\textsubscript{meas}) - \mathcal{F} \left\{ (1 - W) \left(  \mathcal{F}^{-1} [-\textrm{ln}(I\textsubscript{meas})] -  \mathcal{F}^{-1}[\alpha\textsubscript{fit model}] \right) \right\}
\end{equation}

\setcounter{equation}{0}
\setcounter{figure}{0}
\setcounter{table}{0}
\section{NTfit: an E$^{\prime\prime}$-binning algorithm}
 \label{sec:ntfit}

\subsection{Nonuniform Temperature Fitting code (NTfit)}

\begin{figure}[h]
\centering
\includegraphics[width=6.5in]{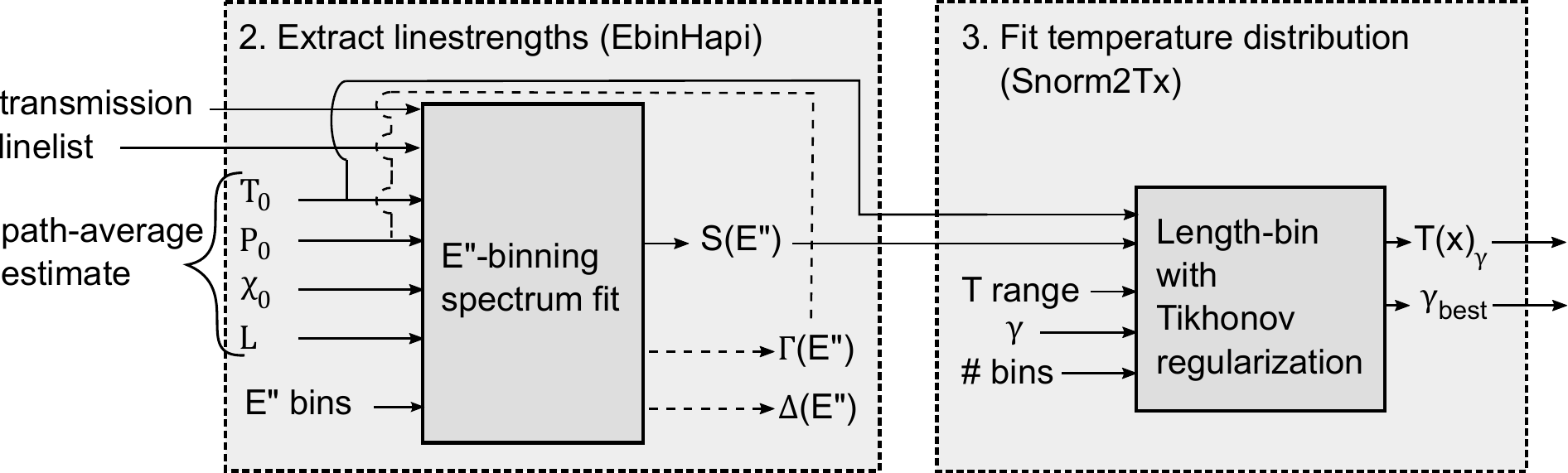}
\caption{Block diagram of 2-step NTfit algorithm. Steps 2 and 3 (same as Fig. \ref{fig:3step}) are saveable classes. Dashed lines indicate optional algorithm steps. EbinHapi and Snorm2Tx are the corresponding class names in the code package.}
\label{fig:blockdiagram}
\end{figure}

The Nonuniform Temperature Fitting (NTfit) code that we developed and used throughout this paper is publicly available in the open-source Python language \cite{ntfit}, which includes example fits.
Figure \ref{fig:blockdiagram} shows the input and output variables for each step of the process.
``Step 2: extract linestrengths,” is the E$^{\prime\prime}$-binning method introduced in Section \ref{sec:ebin}, where one fits the normalized linestrengths $\hat{S}(E^{\prime\prime})$.
As described in Section \ref{sec:lineshape}, one can also fit the Lorentz widths ($\Gamma$) and shifts ($\Delta$) for each E$^{\prime\prime}$-bin, and use this information to update the nominal path-average $T_0$, $P_0$, $\chi_0$.
This optional lineshape-fitting step can produce more accurate normalized linestrengths.
The second block of Fig. \ref{fig:blockdiagram} (``3: fit temperature distribution”), is the length-binning method with Tikhonov regularization described in the companion paper (Section 4 and Appendix D of \cite{malarich_tx1}).
This step takes the set of normalized linestrengths ($\hat{S}(E^{\prime\prime}_i)$) and the reference temperature ($T_0$) from the E$^{\prime\prime}$-binning step.
The rest of this appendix describes the mathematics of E$^{\prime\prime}$-binning as implemented in NTfit (EbinHapi code block in Fig. \ref{fig:blockdiagram}).

Of the potential methods to fit spectra for normalized linestrengths, the NTfit technique is based on the freely available HAPI package \cite{hapi} and HITRAN-formatted linelists, which can fit various variables ($P$, $T$, $\chi$, $L$, frequency-shift $\Delta$).

The two components of a spectral model are the integrated area and the lineshape, both with complex dependence on thermodynamic parameters.
To extract normalized linestrengths, we want to independently adjust both the integrated area and the lineshape to get the lowest fit residual.
NTfit does this spectral fit by splitting one molecular linelist into several different smaller linelists and adjusting the pathlength, pressure, and frequency-shift of each.

The Voigt lineshape profile has four parameters, shown in Table \ref{tab:ntfit}.
The traditional spectroscopic calculations of these parameters are shown under the HAPI definition column.
For E$^{\prime\prime}$-binning we would like to adjust each of the Voigt parameters independently in each E$^{\prime\prime}$-bin, and still work within the HAPI thermodynamic parameter framework, so we add these scaling parameters in the third column of Table \ref{tab:ntfit}.
The HAPI framework can adjust $P$, $T$, $\chi$, $L$, and frequency-shift $\Delta$ for each molecule to fit an absorption spectrum.
But we want the normalized linestrength to maintain a consistent definition over each E$^{\prime\prime}$-bin, particularly with the same reference temperature $T_0$.

\begin{table}[h]
\begin{tabular}{p{3.2cm} p{3.5cm} p{6cm} p{2cm}}
 \toprule
Parameter      &  HAPI definition & NTfit definition   & NTfit free parameter \\
 \midrule 
Integrated area $A$ & $S(T)P \chi L$ & $S(T) P_{act} \frac{P_{scaled}}{P_{act}} \chi L_{path} \frac{L_{scaled}} {L_{path}}$ & $\frac{L_{scaled}}{L_{path}}$ \\
Doppler width & $\nu \sqrt{ \frac {2kT} {mc^2} }$ & same & None \\
Lorentz width $\gamma$ & $P \chi_i \gamma_{296,i} \left(\frac{296}{T} \right)^n$ &  $P_{act} \frac{P_{scaled}}{P_{act}} \chi_i \gamma_{296,i} \left(\frac{296}{T} \right)^n$ & $\frac{P_{scaled}}{P_{act}}$ \\
Linecenter $\nu_0$ & $\nu_0 + P\chi \delta_{296} \left( \frac{296}{T} \right)$ & $\nu_0 + \Delta_{scaled}$ & $\Delta_{scaled}$ \\
 \bottomrule
\end{tabular}
\caption{Voigt lineshape model for NTfit normalized linestrength fitting}
\label{tab:ntfit}
\end{table}

Therefore, for a simplest spectral fit, we only solve for the pathlength ($L_{scaled} / L_{path}$) of each E$^{\prime\prime}$-bin.
This allows us to fit the areas of each E$^{\prime\prime}$-bin without altering the lineshape parameters.
It helps to fit just the normalized linestrengths initially, because it is a linear regression and is thus much faster and more stable than nonlinear regression [45].
The pressure and frequency-shift refinement improves the accuracy, but requires recalculating lineshapes in a much slower, less-stable nonlinear least-squares fit.
In our experience, frequency-shift fitting is particularly unstable, so it may be worth fitting just normalized linestrength and pressures.
The NTfit code allows for manually adjusting the frequency-shift of each E$^{\prime\prime}$-bin, and for selecting which frequency-shifts to fix at manual values in the fit\_snorm\_width\_shift() nonlinear regression.

The HAPI fit parameters may influence multiple Voigt parameters simultaneously. Therefore, the full nonlinear-least-squares fit then reflects upon the normalized linestrength according to Eq. \ref{eq:ntfit}:
\begin{equation}
\label{eq:ntfit}
\hat{S} = \frac {P_{scaled}}{P_{act}} \frac{L_{scaled}}{L_{act}}
\end{equation}

\subsection{Selecting nominal $T_0$, $\chi_0$}
A good estimate of the path-average $T$, $P$, $\chi$, $L$ (in Table \ref{tab:alg}, step 1) approximates the lineshape for normalized linestrength fitting.
In general, we recommend against determining these path-average parameters from a uniform-path fit (unless from a precomputed lookup-table), as it adds unnecessary computation.
We suggest initializing pressure $P_0$ from a barometer or transducer reading, initializing pathlength $L$ from a tape or caliper measurement, and initializing $T_0$ and $\chi_0$ with an educated guess.
Following Table \ref{tab:alg}, step 4, one can update the $T_0$, $\chi_0$ estimates from the initial pathlength-only E$^{\prime\prime}$-bin fit.
This procedure (in NTfit's EbinHapi.fit\_temperature() function) uses an initial fit of normalized linestrengths ( $\hat{S}_{meas,e}$) in Eq. \ref{eq:t0} to estimate a more accurate path-average temperature $T_0$ and molefraction $\chi_0$. 

\begin{equation}
\label{eq:t0}
\min_{T_{fit}, \ \chi_{fit}} \left\{ \sum\limits_{e=1}^{N_{bins}} \left( \hat{S}_{meas,e}\frac{S_e(T_0)\chi_0}{S_e(T_{fit}) \chi_{fit}} - 1  \right)   \right\}
\end{equation}

One can then repeat the pathlength-only fit (using EbinHapi.fit\_snorm()) with the improved $T$, $\chi$ parameters.
Note that Eq. \ref{eq:t0} assumes the pressure and pathlength estimates are correct, so that the molefraction $\chi_{fit}$ absorbs any change to the fitted column density $P\chi L$.
We include a uniform\_temperature\_fit example file in the NTfit package to demonstrate this path-average temperature fit.

As long as the initial pressure $P_0$ and pathlength $L_0$ values are close,
and the (HITRAN) absorption linelist is accurate, this second normalized linestrength fit should produce accurate normalized linestrengths across most E$^{\prime\prime}$ bins.
This entire procedure (Table \ref{tab:alg} through step 4) takes roughly as long as two calculations of the m-FID absorption spectra, or $\sim$10 seconds on our machine.

The power-law correction (Table \ref{tab:alg}, step 5) is a more computationally intensive refinement of this first normalized linestrength estimation,
which typically produces the $<$10\% accuracy improvements to the temperature distribution as shown in Fig. \ref{fig:lineshape}.

\subsection{E$^{\prime\prime}$ estimation}
Another subtlety in NTfit is the lower-state energy calculation.
Step 3 of the algorithm requires a single lower-state energy and normalized linestrength for each E$^{\prime\prime}$-bin.
Yet each E$^{\prime\prime}$-bin is a superposition of many absorption features with different lower-state energies all within some bounding interval of E$^{\prime\prime}$.
To produce a representative E$^{\prime\prime}$, Eq. \ref{eq:e} calculates a sum-of-squares weighted average of all the absorption features in the bin (calculated in the EbinHapi.prep\_ebins() function).
\begin{equation}
\label{eq:e}
E^{\prime\prime}_{out} = \nicefrac {\sum\limits_{i=1}^{lines} S_{T_0,i}^2E^{\prime\prime}_i} {\sum\limits_{i=1}^{lines} S_{T_0,i}^2}
\end{equation}

This weighted average is based on the contribution each absorption feature has to the m-FID signal. The fit uncertainty of the normalized linestrength scales with the 2-norm of the integrated area of each absorption feature in the E$^{\prime\prime}$ bin. It uses the nominal temperature $T_0$ to estimate the linestrength.

This weighted average may not be perfect for a large E$^{\prime\prime}$-bin with a widely-varying normalized linestrength -- a condition most likely to occur in the smallest and largest E$^{\prime\prime}$-bins.
In reality, there may be some horizontal uncertainty bar for the lower-state energy of that E$^{\prime\prime}$-bin; only the vertical uncertainty bar is reflected in the Fig. \ref{fig:3step} temperature distribution fit.
For this reason, you may wish to omit the highest and lowest E$^{\prime\prime}$-bin from the Tikhonov inversion step, particularly if it does not appear to lie on the same curve as the rest of the E$^{\prime\prime}$-bins.

\subsection{Tikhonov regularization function}
The temperature inversion step (Step 3 of Fig. \ref{fig:blockdiagram}) uses Tikhonov regularization, and is described in the companion paper (Section 4 and Appendix D).
We have made a couple updates to the minimization function in this algorithm.
The linestrength residual is weighted by the statistical uncertainties of each E$^{\prime\prime}$-bin.
\begin{equation}
\lVert res \rVert = \sum_{e=1}^{M_{E-bins}} \left(\frac{\hat{S}_{meas, e} - \hat{S}_{fit, e}} {\sigma_{\hat{S}_e}}  \right)
\end{equation}
Where the curve-fit model, $\hat{S}_{fit,e}$ is calculated from a vector $T_i$ of $N$ temperatures and a column-density scaling factor $p\chi L$.

The Tikhonov regularization term, labeled ``temperature curvature" is calculated from the temperature vector $T$.
\begin{equation}
\label{eq:tikh_full}
\lVert reg \rVert = \sum_{i=2}^{N_{length-bins - 1}} (T_i - 1/2 (T_{i-1} + T_{i+1}))^2 \ + (T_N - T_{N-1})^2 = \parallel L \ T \parallel
\end{equation}
We use N = 16 length-bins, and add the extra $(T_N - T_{N-1})$ term to encourage an adiabatic boundary condition at the maximum-temperature $T_N$.
This adiabatic term suggests that the maximum temperature must lie towards the middle of the gas sample, and that the temperature distribution vector $T_i$ suggests the left half of a center-peaked temperature profile.
With the adiabatic term, this Tikhonov regularization term is zero for a uniform-temperature function and near-zero for a left-half convex parabola.
NTfit writes Eq. \ref{eq:tikh_full} in terms of a matrix $L$ in function Snorm2Tx.get\_l2().
One can write the Hessian of this Tikhonov term as $L^T L$ to calculate a Bayesian posterior uncertainty as in \cite{emmert_bayes2020} (see Snorm2Tx.calc\_uc() in NTfit \cite{ntfit})
or to calculate the basis function decomposition of a temperature distribution solution $T_i$.

The uncertainty bars for the furnace measurements were calculated using NTfit's \\ Snorm2Tx.save\_uc\_reg() function.
This procedure used a parameter sweep of $\gamma$ = [1, 0.46, 0.21, 0.1, 0.046, ..., 0.001],
and set corner\_thrsh = .30 to accept retrievals within 30\% of the Euclidean corner described in Appendix D of \cite{malarich_tx1}.

\setcounter{equation}{0}
\setcounter{figure}{0}
\setcounter{table}{0}
\section{Furnace temperature model}
\label{sec:appendix_nc}
\subsection{Natural convection model}

The natural convection model in Figs. \ref{fig:schematic}, \ref{fig:results} from \cite{leong_convection} is a 3D finite-difference solution of the momentum and energy equations (with Boussinesq approximation). The cylinder boundaries on one side are held at temperature $T_H$ and on the other side at $T_C$. Leong \cite{leong_convection} calculates solutions for several geometries, producing 2D contour plots of the non-dimensional temperature ($\theta$ in Eq. \ref{eq:ndtemp}) with respect to position along the vertical centerplane of the tube.

\begin{equation}
\label{eq:ndtemp}
\theta(x,y) = \frac{ T(x,y) - .5*(T_C + T_H) } {T_H - T_C}
\end{equation}

Table \ref{tab:nc} shows the parameters of the closest-matching solution in \cite{leong_convection} compared against our experimental system.
There are two differences in the geometry between the calculations in \cite{leong_convection} and our tube furnace: our tube furnace has a larger aspect ratio (length / diameter), and a longer transition length between the hot-wall and cold-wall portions of the cylinder.
Figure \ref{fig:walltc} adjusts the transition lengths from the value in \cite{leong_convection} to reduce residual with wall-thermocouple data.
These differences do not appear to change the gas temperature profiles, as we can determine a set of boundary conditions that match the laser temperature distributions. The other primary difference in the experimental system is the forced-air flow rate, at 24 and 70 cm\textsuperscript{3}/s. The pipe-flow Reynolds number associated with this follows $Re = u D / \nu$, for inner-diameter $D$ = 8.6 cm, and air kinematic viscosity $\nu \ \sim$ 3e-5 m\textsuperscript{2}/s. So both flow rates are laminar, at Re = 30 for the low-flow case and Re = 90 for the high-flow case.

\begin{table}[h]
\centering
\begin{small}
\centering
\renewcommand\arraystretch{1.3}
\begin{tabular}{ >{\raggedleft}p{2.5cm} cx{4 cm}c cx{5 cm}c cx{2 cm}c}
\toprule
\multicolumn{1}{c}{Name} & Definition & Experiment & Model \\
\midrule
aspect ratio & $L/R$ & 40 & 10 \\
transition length & (distance from $T_H$ to $T_C$)$/L$ & 0.5 & 0.25 \\
b & $z_{T_H} / L $ & 0.5 & 0.5 \\
Rayleigh \# & $ \left(g R^3 \left(T_H - T_C \right)\right)/\left(\nu \alpha \overline{T}\right)$ & 200,000 & 20,000 \\
velocity$_{NC}$ & $\sqrt{g R (T_H - T_C)/\overline{T}}$ & 60 cm/s & \\
velocity$_{forced \ air}$ & Flow rate / Area\textsubscript{circle} & 1 cm/s & 0 \\
radiation & & semi-transparent walls & none \\
hot end-cap & boundary condition & $Q=$0 center of tube & $T_H$ \\
\bottomrule
\end{tabular}
\end{small}
\caption{Natural convection model parameters from \cite{leong_convection}.}
\label{tab:nc}
\end{table}

\subsection{Temperature bias in thermocouple rod}
The thermocouple rod measurements in Fig. \ref{fig:schematic} record some intermediate temperature between the true gas temperature and nearby wall temperaturesdue to the complex interplay of convection and radiation around the thermocouple bead.
The measurements are collected with a bare-bead thermocouple, and there is no radiative correction applied to the recorded temperature.
The radiative transfer is affected by the view factor between the thermocouple bead and the cylinder walls.
The view factor between a centered thermocouple bead and the tube walls within an axial distance $d$ is given by Eq. \ref{eq:vf}
\begin{equation}
VF(d) = \left(1 + \left(\frac{r_{tube}}{d}\right)^2   \right)^{-1/2}
\label{eq:vf}
\end{equation}
For this 10 cm ID tube ($r_{tube}$ = 5 cm), Eq. \ref{eq:vf} estimates a 70\% view factor between the bead and the the tube walls within 5 cm of the bead, and a 90\% view factor between the bead and the tube walls within 10 cm.
In Fig. \ref{fig:schematic}, thermocouple rod measurements around x=10 cm are more than 100 K above all of the wall-thermocouple measurements within 10 cm (a 90\% view factor).
The blurring of the axial wall temperature gradients due to a radiative balance over different view factors cannot explain the magnitude of this discrepancy between the wall and rod thermocouples.
Therefore, we believe that natural convection cells are changing the thermocouple rod temperature. 

Furthermore, it is challenging to correct this thermocouple rod measurement without knowledge of the natural convection model, due to the unknown relative radiation and convection losses from the thermocouple bead.
Radiation corrections require some estimate of system parameters such as fluid velocity \cite{thermocouple_rad_correct}, which are not known for this system.
Use of a shielded-tip thermocouple, rather than the bare-bead thermocouple used in this paper, would reduce the radiation bias on the temperature reading \cite{shielded_tc}. We recommend using a shielded thermocouple for future demonstration experiments.

\end{document}